\documentclass[a4paper,11pt]{article}
\pdfoutput=1
\usepackage{./auxi/jheppub}
\usepackage[utf8]{inputenc}

\usepackage{slashed} 
\usepackage{mathtools} 
\usepackage{tikz}
\usepackage{dsfont} 
\usetikzlibrary{positioning,arrows,calc}
\usetikzlibrary{arrows.meta}
\usetikzlibrary{decorations.pathmorphing}
\usetikzlibrary{decorations.markings}
\usetikzlibrary{shapes.geometric}
\usetikzlibrary{patterns}
\usepackage{xparse}
\usepackage{diagbox} 
\usepackage{tensor} 
\usepackage{subcaption} 

\DeclareMathOperator{\Disc}{Disc}
\DeclareMathOperator{\dDisc}{dDisc}
\DeclareMathOperator{\Res}{Res}

\newcommand{\vvev}[1]{\langle\!\langle\, #1 \, \rangle\!\rangle}

\newcommand{\Dh}{\hat{\Delta}}

\newcommand{\Wl}{\mathcal{W}_\ell}

\newcommand{\fh}{\hat{f}}

\def\Fm{{\mathcal{F}}}
\def\Gm{{\mathcal{G}}}

\def\Nm{{\mathcal{N}}}
\def\Om{{\mathcal{O}}}
\def\Pm{{\mathcal{P}}}

\def\a{{\alpha}}

\usepackage{bbm}

\def\veps{\varepsilon}

\newcommand\zb{{\bar{z}}}

\newcommand\Dp{{\Delta_\phi}}

\def\wh{\widehat}

\newif\ifstartcompletesineup
\newif\ifendcompletesineup
\pgfkeys{
    /pgf/decoration/.cd,
    start up/.is if=startcompletesineup,
    start up=true,
    start up/.default=true,
    start down/.style={/pgf/decoration/start up=false},
    end up/.is if=endcompletesineup,
    end up=true,
    end up/.default=true,
    end down/.style={/pgf/decoration/end up=false}
}
\pgfdeclaredecoration{complete sines}{initial}
{
    \state{initial}[
        width=+0pt,
        next state=upsine,
        persistent precomputation={
            \ifstartcompletesineup
                \pgfkeys{/pgf/decoration automaton/next state=upsine}
                \ifendcompletesineup
                    \pgfmathsetmacro\matchinglength{
                        0.5*\pgfdecoratedinputsegmentlength / (ceil(0.5* \pgfdecoratedinputsegmentlength / \pgfdecorationsegmentlength) )
                    }
                \else
                    \pgfmathsetmacro\matchinglength{
                        0.5 * \pgfdecoratedinputsegmentlength / (ceil(0.5 * \pgfdecoratedinputsegmentlength / \pgfdecorationsegmentlength ) - 0.499)
                    }
                \fi
            \else
                \pgfkeys{/pgf/decoration automaton/next state=downsine}
                \ifendcompletesineup
                    \pgfmathsetmacro\matchinglength{
                        0.5* \pgfdecoratedinputsegmentlength / (ceil(0.5 * \pgfdecoratedinputsegmentlength / \pgfdecorationsegmentlength ) - 0.4999)
                    }
                \else
                    \pgfmathsetmacro\matchinglength{
                        0.5 * \pgfdecoratedinputsegmentlength / (ceil(0.5 * \pgfdecoratedinputsegmentlength / \pgfdecorationsegmentlength ) )
                    }
                \fi
            \fi
            \setlength{\pgfdecorationsegmentlength}{\matchinglength pt}
        }] {}
    \state{downsine}[width=\pgfdecorationsegmentlength,next state=upsine]{
        \pgfpathsine{\pgfpoint{0.5\pgfdecorationsegmentlength}{0.5\pgfdecorationsegmentamplitude}}
        \pgfpathcosine{\pgfpoint{0.5\pgfdecorationsegmentlength}{-0.5\pgfdecorationsegmentamplitude}}
    }
    \state{upsine}[width=\pgfdecorationsegmentlength,next state=downsine]{
        \pgfpathsine{\pgfpoint{0.5\pgfdecorationsegmentlength}{-0.5\pgfdecorationsegmentamplitude}}
        \pgfpathcosine{\pgfpoint{0.5\pgfdecorationsegmentlength}{0.5\pgfdecorationsegmentamplitude}}
}
    \state{final}{}
}

\definecolor{bgbox}{RGB}{255,254,230}
\definecolor{setupplane}{RGB}{230,230,230}
\definecolor{gluoncolor}{RGB}{207,54,108}
\definecolor{vertexcolor}{RGB}{53,152,219}
\definecolor{SEcolor}{RGB}{176,156,255}
\definecolor{blobcolor}{RGB}{190,180,230}
\tikzset{
corner/.style={line width=1pt,dashed,draw=black,dash pattern=on 6pt off 4pt},
scalar/.style={line width=1pt,draw=black},
gluon/.style={line width=1pt,decorate, draw=gluoncolor,
    decoration={complete sines,aspect=0,amplitude=1.25mm,segment length=1.5mm,start up,end up}},
ghost/.style={line width=1pt,loosely dotted,draw=black},
wilson/.style={line width=2pt,draw=black},
 }
\NewDocumentCommand\semiloop{O{black}mmmO{}O{above}}
{%
\draw[#1] let \p1 = ($(#3)-(#2)$) in (#3) arc (#4:({#4+180}):({0.5*veclen(\x1,\y1)})node[midway, #6] {#5};)
}

\usepackage{simplewick}
\usepackage{dsfont}
\usepackage{float}
\usepackage{pgfplots}
\pgfplotsset{compat=1.14}

\usepackage{tikz}
\usetikzlibrary{shapes}
\usetikzlibrary{decorations.pathmorphing}
\usetikzlibrary{decorations.markings}
\tikzset{
  snake it/.style={
    decorate, 
    decoration=snake,
    segment length=3
  }
}

\let\oldbfseries=\bfseries
\let\oldmdseries=\mdseries
\let\oldnormalfont=\normalfont
\renewcommand{\bfseries}{\oldbfseries\boldmath}
\renewcommand{\mdseries}{\oldmdseries\unboldmath}
\renewcommand{\normalfont}{\oldnormalfont\unboldmath}

\makeatletter
\newlength{\apb@width}
\newcommand{\autoparbox}[2][c]{\settowidth{\apb@width}{#2}\parbox[#1]{\apb@width}{#2}}

\makeatother

\DeclareMathOperator{\tr}{tr}

\def\Fm{{\mathcal{F}}}
\def\Gm{{\mathcal{G}}}

\def\Nm{{\mathcal{N}}}
\def\Om{{\mathcal{O}}}
\def\Pm{{\mathcal{P}}}

\def\zb{{\bar{z}}}

\def\a{{\alpha}}

\def\veps{\varepsilon}

\newcommand{\beq}{\begin{equation}}
\newcommand{\eeq}{\end{equation}}

\definecolor{nicegreen}{rgb}{0.1,0.6,0.1}

\newmuskip\pFqskip
\mathchardef\pFcomma=\mathcode`,

\makeatletter
\renewcommand*\env@matrix[1][\arraystretch]{%
  \edef\arraystretch{#1}%
  \hskip -\arraycolsep
  \let\@ifnextchar\new@ifnextchar
  \array{*\c@MaxMatrixCols c}}
\makeatother

\graphicspath{{./figures/}}

\allowdisplaybreaks 

\title{\center{A dispersion relation for defect CFT}}

\author[1]{Julien Barrat,}
\author[2]{Aleix Gimenez-Grau,}
\author[2]{Pedro Liendo.}

\affiliation[1]{Institut für Physik und IRIS Adlershof, Humboldt-Universität zu Berlin,
 Zum Großen Windkanal 2, 12489 Berlin, Germany}
\affiliation[2]{DESY Hamburg, Theory Group, Notkestra{\ss}e 85, D-22607 Hamburg, Germany}

\emailAdd{julien.barrat@hu-berlin.de, aleix.gimenez@desy.de, pedro.liendo@desy.de}

\preprint{HU-EP-22/18-RTG}

\abstract{We present a dispersion relation for defect CFT that reconstructs two-point functions in the presence of a defect as an integral of a single discontinuity.
	The main virtue of this formula is that it streamlines explicit bootstrap calculations, bypassing the resummation of conformal blocks.
	As applications we reproduce known results for monodromy defects in the epsilon-expansion, and present new results for the supersymmetric Wilson line at strong coupling in $\Nm=4$ SYM. In particular, we derive a new analytic formula for the highest $R$-symmetry channel of single-trace operators of arbitrary length.	
	}

\begin{document} 
	
\maketitle

\newpage

\flushbottom

\section{Introduction}
\label{sec:intro}

The modern analytic bootstrap is a powerful approach to study the dynamics of conformal field theories (CFTs) starting from basic principles, such as unitarity and crossing. The flagship result of this line of thinking is arguably the Lorentzian inversion formula \cite{Caron-Huot:2017vep}, which reconstructs the CFT data starting from the double discontinuity $\dDisc \Gm$ of the CFT correlator. This is a powerful calculational tool, as the double discontinuity is a simpler object that can often be approximated numerically \cite{Albayrak:2019gnz,Liu:2020tpf,Atanasov:2022bpi} or computed order by order in perturbation theory \cite{Alday:2017vkk,Alday:2017zzv,Alday:2019clp}.

Building upon this result, the authors of \cite{Carmi:2019cub} went one step ahead and wrote a dispersion relation that reconstructs the full correlator (as opposed to the CFT data) starting from $\dDisc \Gm$. This equation is reminiscent of the dispersion relations familiar from S-matrix theory. The logic of the derivation is simple: starting from the inversion formula, one extracts the CFT data in terms of $\dDisc \Gm$, which is then plugged in the block expansion. After performing the infinite sums, the final result is an integral over $\dDisc \Gm$ times a theory-independent kernel. The main technical achievement of \cite{Carmi:2019cub} is a closed-form expression for the integration kernel. 
The position-space dispersion relation is a central piece in our understanding of analytic bootstrap: on one hand, it is equivalent to the dispersion relation in Mellin space \cite{Penedones:2019tng,Carmi:2020ekr}; on the other hand, it can be used to obtain the analytic functionals of \cite{Mazac:2019shk}; finally, it leads to dispersive sum rules equivalent to the crossing-symmetric Mellin bootstrap \cite{Gopakumar:2016cpb,Gopakumar:2021dvg}.
This shows that the main approaches to the analytical bootstrap are ultimately equivalent \cite{Caron-Huot:2020adz}.

In parallel to the analytic bootstrap for four-point CFT correlators, significant progress has been made on adapting and generalizing bootstrap techniques to include extended objects or defects. The setup that has gotten the most attention is two-point functions of local operators in the presence of a defect. The kinematics of this configuration is similar to the four-point function case, because correlators depend on two conformal invariants and satisfy a ``crossing equation'' that equates two OPE channels. In the defect setup the name ``crossing'' is not quite accurate, as the channels are different in nature: there is a \textit{defect channel} OPE that expands a local operators as an infinite sum of defect excitations, and a more familiar \textit{bulk channel} where the two local operators are fused together. Motivated by the results of \cite{Carmi:2019cub}, in this work we present a dispersion relation for defect CFT, which reconstructs two-point functions starting from a single discontinuity  of the correlator:\footnote{We should stress that we are using the term ``dispersion relation'' loosely. The discontinuity of defect CFT correlators does not have well-defined positivity properties \cite{Lemos:2017vnx}, and cannot be interpreted as the absorptive part of a physical process.}
\begin{align}
\setlength\fboxsep{.7em}
 \boxed{
 F(r,w)
 = \int_0^r \frac{dw'}{2 \pi i}
 \frac{w (1 - w') (1 + w')}{ w' (w'-w) (1 - w w')}
 \Disc F(r,w') \, . }
 \label{eq:disp-rel}
\end{align}
Here $F(r,w)$ is a two-point function in the presence of a defect, while $r$ and $w$ are two conformal cross-ratios to be defined below.
As already pointed out, the advantage of working with discontinuities is that they are simpler objects, that can often be computed even when the full correlator is not known. Furthermore, our formula can be used to derive the correlator itself, bypassing the resummation of the CFT data. Once the correlator is obtained, the CFT data can still be extracted by means of a block expansion, a process that is usually simpler. The presence of a single discontinuity in our result can be traced back to the inversion formula presented in \cite{Lemos:2017vnx}, which reconstructs the defect data starting from $\Disc F$. 

We should point out that in defect CFT there is a second inversion formula that reconstructs the \textit{bulk data} starting from a double discontinuity $\dDisc F$. 
In \cite{Liendo:2019jpu}, this second inversion formula was shown to be equivalent to the original four-point function inversion formula \cite{Caron-Huot:2017vep}, provided one identifies certain kinematical parameters between the two configurations.
In principle, one could write a defect CFT dispersion relation that reconstructs the correlator starting from $\dDisc F$, but we expect this to be equal to the dispersion relation in \cite{Carmi:2019cub,Trinh:2021mll} with suitable identification of the parameters.

The organization of the paper is as follows. We start in section \ref{sec:projective} studying the closely related setup of projective space: in this configuration two-point functions depend on a single cross-ratio, so the kinematics is somewhat simpler. After this warm-up example, we derive our main result in section \ref{sec:dispersion}, a dispersion relation for any defect CFT with codimension two or higher. In section \ref{sec:monodromy} we use our formula to rederive known results for monodromy defects in the $\veps$--expansion. These examples serve as a consistency check, but also to demonstrate the simplicity of using the dispersion relation in explicit calculations.
In section \ref{sec:newresults} we turn our focus to the supersymmetric Wilson line in $\Nm=4$ SYM. Here we extend the methods of \cite{Barrat:2021yvp} and derive new correlators for the case when the external half-BPS operators have unequal scaling dimensions. 
Thanks to the dispersion relation, the calculations simplify greatly and we manage to obtain a closed-form expression for the highest $R$-symmetry channel in a correlator of arbitrary half-BPS operators.
We finish with a discussion of possible future directions in section \ref{sec:conclusions} and relegate some technical results to an appendix. 

\paragraph{Note added:} While this paper was in preparation, we became aware of \cite{bianchibonomi}, whose content partially overlaps with the present work. We coordinated with the authors for a simultaneous submission.

\section{Warm-up example: projective space}
\label{sec:projective}

Before we jump into the study of general conformal defects, we take a short detour and consider CFT in projective space \cite{Nakayama:2016xvw,Nakayama:2016cim,Hasegawa:2016piv,Hogervorst:2017kbj,Hasegawa:2018yqg,Giombi:2020xah}.
The kinematics of CFT in projective space are similar to boundary and defect CFT, and in particular, correlators satisfy a crossing equation that can be studied with bootstrap methods.
As a result, it is the perfect toy model to motivate the techniques we use in the upcoming sections.
After reviewing basic properties of the setup, we bootstrap two-point functions in the Wilson-Fisher fixed point to order $O(\veps^2)$ with a dispersion relation derived in \cite{Giombi:2020xah}.
Although the results are not new \cite{Hasegawa:2018yqg,Giombi:2020xah}, our methods are simpler than previous approaches.
We conclude this section with some comments on the possibility of applying similar methods to boundary CFT.

\subsection{Review of projective space CFT}
\label{subsec:projective-setup}

Let us start with a quick review of CFT in projective space.
We follow the conventions of \cite{Giombi:2020xah}, where the reader can also find a more thorough discussion.
Real projective space $\mathbb{RP}^d$ is obtained from Euclidean space $\mathbb{R}^d$ by identifying points under inversion
\begin{align}
 x^\mu \to - \frac{x^\mu}{x^2} \, .
\end{align}
Local operators in $\mathbb{RP}^d$ transform under inversion as $\Om \to \pm \Om$, where the sign is the $\mathbb{Z}_2$ charge under inversion of the operator.
When considering correlation functions, the overall charge has to vanish in order to obtain a non-zero result.
In particular, an even scalar operator can acquire a one-point function 
\begin{align}
 \langle \Om(x) \rangle
 = \frac{a_\Om}{(1 + x^2)^\Delta} \, ,
 \label{eq:one-pt-def}
\end{align}
while spinning operators and odd scalars have vanishing one-point function.
The two-point functions of scalars has a similar prefactor, although it depends on a function of a cross-ratio $\eta$:
\begin{align}
 \langle \Om_1(x_1) \Om_2(x_2) \rangle_{\mathbb{RP}^d}
 = \frac{F(\eta)}{(1 + x_1^2)^{\Delta_1} (1 + x_2^2)^{\Delta_2}} \, , \qquad
 \eta = \frac{(x_1 - x_2)^2}{(1 + x_1^2)(1 + x_2^2)} \, .
 \label{eq:two-pt-rpd}
\end{align}
We follow standard conventions in the CFT literature, where the two-point function is unit-normalized when the two operators approach each other
\begin{align}
 \lim_{x_1 \to x_2} \langle \Om_1(x_1) \Om_2(x_2) \rangle
 \approx \frac{\delta_{\Om_1\Om_2}}{(x_1 - x_2)^{2\Delta_1}} \, .
\end{align}
As a result, the one-point coefficient $a_\Om$ defined in \eqref{eq:one-pt-def} cannot be rescaled away, and it contains dynamical information of the CFT.

The two-point correlator $F(\eta)$ satisfies a crossing equation, which is obtained comparing \eqref{eq:two-pt-rpd} to the case when one operator is replaced by its image under inversion.
Under this transformation one finds $\eta \to 1-\eta$, and including a sign from the charge of $\Om_1$ and $\Om_2$ under inversion, the crossing equation reads
\begin{align}
 F(\eta) = \pm F(1-\eta) \, .
\end{align}
As usual in CFT, the product of two local operators can be expanded in a convergent sum over local operators, the so called operator product expansion (OPE).
For two-point functions, the OPE implies that $F(\eta)$ admits a conformal block decomposition, and equating the direct- and the inverse-channel expansions gives the crossing equation
\begin{align}
 & F(\eta)
 = \sum_\Om \mu_{\Om_1\Om_2\Om} f_{\Delta}(\eta)
 = \pm \sum_\Om \mu_{\Om_1\Om_2\Om} f_{\Delta}(1-\eta) \, .
 \label{eq:crossing-rpd}
\end{align}
The expansion coefficients are the product $\mu_{\Om_1\Om_2\Om} \equiv \lambda_{\Om_1\Om_2\Om} a_\Om$ of a three-point OPE coefficient $\lambda_{\Om_1\Om_2\Om}$ and a one-point function $a_\Om$.
Note that because of \eqref{eq:one-pt-def}, the sum runs only over $\mathbb{Z}_2$-even scalar operators.
The conformal blocks are responsible of summing contributions of descendants, such that the sum in \eqref{eq:crossing-rpd} runs only over primaries.
The easiest way to determine the blocks is by solving a Casimir equation, giving \cite{Nakayama:2016xvw}
\begin{align}
 f_{\Delta}(\eta) 
 = \eta^{\frac{\Delta - \Delta_1 - \Delta_2}{2}} {}_2F_1 \left( 
    \frac{\Delta + \Delta_{12}}{2} ,
    \frac{\Delta - \Delta_{12}}{2} ,
    \Delta + 1 - \frac{d}{2} ;
    \eta
 \right) \, .
\end{align}

Finally, let us derive a dispersion relation for the two-point correlator $F(\eta)$, which first appeared in \cite{Giombi:2020xah}.
The derivation relies on the analyticity properties of $F(\eta)$ in the complex plane.
It is well-known that in Lorentzian signature the correlator has branch-cut singularities when two operators become light-like separated.
In our setup there are two possibilities, namely that $\Om_1$ approaches the lightcone of $\Om_2$, and that $\Om_1$ approaches the lightcone of the image of $\Om_2$, which lead to two branch cuts $\eta \in (-\infty, 0]$ and $\eta \in [1, \infty)$.
Keeping this in mind, we can start from Cauchy's integral formula and deform the contour picking contributions along the two cuts.
The resulting formula is the desired dispersion relation
\begin{align}
 & F(\eta)
 = \int_{-\infty}^{0} \frac{d\zeta}{2\pi i} \frac{\Disc F(\zeta)}{\zeta - \eta}
 + \int_{1}^{\infty}   \frac{d\zeta}{2\pi i} \frac{\Disc F(\zeta)}{\zeta - \eta} \, ,
 \label{eq:disp-rel-rpd} \\[0.5em]
 & \Disc F(\zeta)
 = F(\zeta + i 0) - F(\zeta - i 0) \, ,
\end{align}
which was already derived in \cite{Giombi:2020xah}.
For a single direct-channel block, the discontinuity along the $\zeta \in (-\infty, 0]$ branch cut reads
\begin{align}
 \Disc_{\zeta<0} f_{\Delta}(\zeta)
 & = 2 i \sin \left(\frac{\pi  (\Delta -\Delta_1 - \Delta_2)}{2}\right)
 (-\zeta)^{\frac{\Delta - \Delta_1 - \Delta_2}{2}} \notag \\
 & \qquad \times {}_2F_1 \left( 
    \frac{\Delta + \Delta_{12}}{2} ,
    \frac{\Delta - \Delta_{12}}{2} ,
    \Delta + 1 - \frac{d}{2} ;
    \zeta
 \right) \, ,
 \label{eq:disc-rpd}
\end{align}
while a similar formula holds for the discontinuity along $\zeta \in [1, +\infty)$ with inverted-channel blocks.
The most important property of \eqref{eq:disc-rpd} is that for operators near the multi-twist dimension $\Delta = \Delta_1 + \Delta_2 + 2n + \gamma$,  the discontinuity is proportional to the anomalous dimension $\Disc \sim \gamma$.
In what follows, we show that this property allows to bootstrap the Wilson-Fisher correlator in an effortless manner.

\subsection{Wilson-Fisher fixed point}
\label{subsec:projective-wf}

In order to show the power of dispersion relations, we focus our attention to the Wilson-Fisher fixed point, which can be studied perturbatively in powers of $\veps=4-d$.
Here we consider the two-point function $\langle \phi^i(x_1) \phi^j(x_2) \rangle$ of the fundamental scalar $\phi^i(x)$, which transforms as a vector under $O(N)$.
By using the dispersion relation \eqref{eq:disp-rel-rpd}, we can fully reconstruct the correlator to order $O(\veps^2)$ in a simple fashion.

Let us start analyzing the structure of the OPE $\phi \times \phi$, which is well understood thanks to recent progress on the conformal bootstrap, see \cite{Henriksson:2022rnm} for a comprehensive review.
For us it is important that the three-point OPE coefficients are of order \cite{Alday:2017zzv,Henriksson:2018myn}
\begin{align}
\lambda_{\phi\phi[\phi\phi]_{\ell,0}}^2 \sim O(1) \, , \qquad 
\lambda_{\phi\phi[\phi\phi]_{\ell,1}}^2 \sim O(\veps^2) \, , \qquad
\lambda_{\phi\phi[\phi\phi]_{\ell,n\ge2}}^2 \sim O(\veps^3) \, ,
\end{align}
where $[\phi\phi]_{\ell,n}$ is an operator with approximate dimension $\Delta_{\ell,n} = 2\Dp + \ell + 2n + \gamma_{\ell,n}$.
In the case of projective CFT, only scalars $\ell=0$ survive in the OPE, which we label with $n$.
Remember that in the block expansion each contribution is proportional to $\mu_{\phi\phi\Om} = \lambda_{\phi\phi\Om} a_\Om$, so to order $O(\veps^2)$ we find
\begin{align}
  F(\eta)
& = f_{1}
  + \big( \mu_{\phi\phi0}^{(0)} + \veps \mu_{\phi\phi0}^{(1)} + \ldots \big) 
    f_{2\Delta_\phi + \veps \gamma_0^{(1)} + \veps^2 \gamma_0^{(2)} + \ldots} \notag \\
& + \big( \mu_{\phi\phi1}^{(1)} \veps + \ldots \big) 
    f_{2\Delta_\phi + 2 + \veps \gamma_1^{(1)} + \ldots}
  + \sum_{n = 2}^\infty \big( \mu_{\phi\phi n}^{(2)} \veps^2 + \ldots \big) 
    f_{2\Delta_\phi + 2n + \ldots} \, .
 \label{eq:structure-ope}
\end{align}
The first operator is the identity, the second operator is $[\phi\phi]_{0,0} = \phi^2$, and the third operator is $[\phi\phi]_{0,1} = \phi^4$.
The anomalous dimensions of $\phi$, $\phi^2$ and $\phi^4$ are well known
\begin{align}
 \gamma_\phi^{(2)} = \frac{N+2}{4(N+8)^2} \, , \quad
 \gamma_0^{(1)} = \frac{N+2}{N+8} \, , \quad
 \gamma_0^{(2)} = \frac{6(N+2)(N+3)}{(N+8)^3} \, , \quad
 \gamma_1^{(1)} = 1 \, 
 \label{eq:cft-data-wf}.
\end{align}
We shall treat these as input, which can be obtained from the bootstrap of four-point functions \cite{Gopakumar:2016cpb,Dey:2016mcs,Alday:2017zzv,Henriksson:2018myn}, or from Feynman diagram calculations \cite{Kleinert:1991rg}.
Our goal is to extract the information which is intrinsic to projective space CFT, namely the coefficients $\mu_{\phi\phi n}$.

The starting point of our bootstrap calculation is to compute the discontinuity.
Remember that from \eqref{eq:disc-rpd} we conclude that $\mu_{\phi\phi n} \Disc f_{2\Dp+2n+\gamma_n} \propto \mu_{\phi\phi n} \gamma_n$. 
As a result, the infinite sum in \eqref{eq:structure-ope} starts to contribute to $\Disc$ at order $O(\veps^3)$.
In other words, if we restrict ourselves to $O(\veps^2)$ the discontinuity can be computed from only three operators:
\begin{align}
  \Disc F(\eta) \big|_{O(\veps^2)}
& = \Disc \Big( 
    f_{1}
  + \mu_{\phi\phi0} f_{2\Delta_0} 
  + \mu_{\phi\phi1} f_{2\Delta_1} 
  \Big) \, .
 \label{eq:structure-disc}
\end{align}
For each of these operators we can use the general form of the discontinuity \eqref{eq:disc-rpd} and expand the result to order $O(\veps^2)$:
\begin{align}
  \frac{1}{2\pi i}
  \underset{\zeta<0}{\Disc} \, F(\zeta)
& = - \frac{\sin (\pi  \Delta_\phi)}{\pi (-\zeta )^{\Delta_\phi}}
  + \frac{\gamma_0^{(1)} \mu_{\phi\phi0}^{(0)} \veps
         + \big(
             \gamma_0^{(2)} \mu_{\phi\phi0}^{(0)}
         +   \gamma_0^{(1)} \mu_{\phi\phi0}^{(1)}
         + 2 \gamma_1^{(1)} \mu_{\phi\phi1}^{(1)} \big) \veps^2 }{
           2 (1-\zeta)} \notag \\
& \quad
  + (\gamma_0^{(1)})^2 \mu_{\phi\phi0}^{(0)} \veps^2 \frac{\log (-\zeta )}{4 (1-\zeta)}
  + \log (1-\zeta ) \left(
          \frac{\gamma_1^{(1)} \mu_{\phi\phi1}^{(1)}}{\zeta }
        + \frac{\gamma_0^{(1)} \mu_{\phi\phi0}^{(0)}}{4 (1-\zeta)}
     \right) \veps^2  \, .
\end{align}
Since we have the discontinuity, it is a simple exercise to reconstruct the correlator using the dispersion relation \eqref{eq:disp-rel-rpd}.
All one has to do is to compute several elementary integrals, giving
\begin{align}
  F(\eta) 
& = \frac{1}{\eta^{\Delta_\phi}}
  + \left( 
        \gamma_0^{(1)} \mu_{\phi\phi0}^{(0)} \veps + \big(
        \gamma_0^{(2)} \mu_{\phi\phi0}^{(0)}
    +   \gamma_0^{(1)} \mu_{\phi\phi0}^{(1)}
    + 2 \gamma_1^{(1)} \mu_{\phi\phi1}^{(1)} \big) \veps^2
    \right) \frac{\log \eta}{2(1-\eta)} 
  \notag \\
& \quad
  + (\gamma_0^{(1)})^2 \mu_{\phi\phi0}^{(0)} \veps^2 \frac{\log^2 \eta}{8 (1-\eta)}
  + \gamma_1^{(1)} \mu_{\phi\phi1}^{(1)} \veps^2 \frac{\text{Li}_2\, \eta + \log (1-\eta ) \log \eta}{\eta }
  \notag \\
& \quad
  + \frac14 \gamma_0^{(1)} \mu_{\phi\phi0}^{(0)} \veps^2 
    \frac{\text{Li}_2 \, \eta + \log (1-\eta ) \log \eta - \pi ^2/6}{1-\eta}
  \pm \big( \eta \to 1-\eta \big)\, .
  \label{eq:RP-full-corr}
\end{align}
This result still depends on three unknown coefficients $\mu_{\phi\phi0}^{(0)}$, $\mu_{\phi\phi0}^{(1)}$ and $\mu_{\phi\phi1}^{(0)}$.
However, these coefficients can be fixed expanding the correlator \eqref{eq:RP-full-corr} in conformal blocks, and demanding the result is consistent with the definitions \eqref{eq:structure-ope}.
The solution to the resulting equations is
\begin{align}
 \mu_{\phi\phi0}^{(0)} = \pm 1 \, , \qquad 
 \mu_{\phi\phi0}^{(1)} = - \frac{N+2}{2 (N+8)} \, , \qquad 
 \mu_{\phi\phi1}^{(1)} = \frac{N+2}{4 (N+8)} \, .
 \label{eq:cft-data-rpd}
\end{align}
Furthermore, it is straightforward to also extract the OPE coefficients $\mu_{\phi\phi n}^{(2)}$.
Our final result is the combination of \eqref{eq:RP-full-corr} with \eqref{eq:cft-data-rpd} and \eqref{eq:cft-data-wf}, which is equal to the correlator obtained in \cite{Giombi:2020xah}.
Besides the simplicity of the calculations, one advantage of our method is that, unlike the equations of motion method of \cite{Giombi:2020xah}, it does not rely on a particular Lagrangian description.

\subsection{Comments on boundary CFT}
\label{subsec:bcft}

It is well known that projective space CFT shares many features with boundary CFT (BCFT), see table 1 of \cite{Giombi:2020xah} for a nice illustration.
It is then natural to wonder if methods similar to the above can be used to bootstrap two-point functions in BCFT.\footnote{The bootstrap approach to BCFT is a large subject, and it is hard to do it justice here. Some works similar to the present paper are \cite{Liendo:2012hy,Bissi:2018mcq,Mazac:2018biw,Kaviraj:2018tfd,Gimenez-Grau:2020jvf,Dey:2020jlc}, where methods to solve crossing analytically have been used, but see also \cite{Cuomo:2021cnb,Padayasi:2021sik,DiPietro:2020fya,Behan:2020nsf,Behan:2021tcn} for an incomplete list of recent work.}
The BCFT two-point function of identical scalars may be expressed as
\begin{align}
 \langle \Om(x_1) \Om(x_2) \rangle_{\text{BCFT}}
 = \frac{\eta^{\Delta}}{(x_1^\bot x_2^\bot)^{\Delta}}
   F(\eta) \, , \qquad
 \eta = \frac{4 x_1^\bot x_2^\bot}{(x_{12}^{\|})^2 + (x_1^\bot + x_2^\bot)^2} \, ,
\end{align}
where $x^\bot \ge 0$ is a coordinate orthogonal to the boundary, and $x^{\|}$ are coordinates along the boundary.
The correlator $F(\eta)$ has two branch cuts $\eta \in (-\infty, 0]$ and $\eta \in [1, \infty)$, so naively it can be reconstructed from the dispersion relation \eqref{eq:disp-rel-rpd}.
However, it can be shown that the correlator does not decay as $|\eta| \to \infty$, and instead it is bounded by a constant \cite{Mazac:2018biw,Kaviraj:2018tfd}
\begin{align}
 |F(\eta)| \le K \quad \text{as} \quad |\eta| \to \infty \, .
\end{align}
This means that in deriving \eqref{eq:disp-rel-rpd}, it is not possible to drop the arcs at infinity, so the dispersion relation is invalid.
Possibly this issue can be fixed by considering a subtracted version of \eqref{eq:disp-rel-rpd}, but we shall not pursue this here.

Besides this, a second and more problematic issue is that we expect there should exist a dispersion relation more powerful than \eqref{eq:disp-rel-rpd}.
Indeed it was shown in \cite{Mazac:2018biw,Kaviraj:2018tfd} that there exist analytic functionals with double zeros when acting on boundary blocks, and single zeros acting on bulk blocks.
This suggest the existence of a dispersion relation that reconstructs the correlator from a double discontinuity in the boundary channel, and a single discontinuity in the bulk channel:
\begin{align}
 F(\eta) \sim 
   \int K_{\text{bdy}}(\eta) \dDisc F(\eta)
 + \int K_{\text{blk}}(\eta) \Disc F(\eta) \, .
\end{align}
Obtaining such a formula would be a very interesting result, that would simplify future analytic studies of BCFT.

\section{Dispersion relation}
\label{sec:dispersion}

In this section we present the dispersion relation for defect CFT. 
Given the importance of this result, we derive it in two different ways: first from Cauchy's integral formula, and second using the Lorentzian inversion formula. 
We also study the special case of codimension-two defects, and we conclude reviewing how to use subtractions when the correlator does not decay fast enough at infinity.

\subsection{Main result}
\label{subsec:mainresults}

The main object in this work is the two-point function of bulk scalars $\phi_i$ in the presence of a conformal defect:
\begin{align}
\label{eq:two-pt-with-def}
 \vvev{ \phi_1(x_1) \, \phi_2(x_2) }  
 = \frac{F(r,w)}{|x_1^\bot|^{\Delta_1} |x_2^\bot|^{\Delta_2}} \, .
\end{align}
The double-bracket notation is a reminder that the expectation value is taken in the presence of the conformal defect.
We use the standard definition of conformal defect, namely it is an extended object that preserves the subgroup
\begin{align}
 SO(p+1,1) \oplus SO(q) \subset SO(d+1,1) \, , \qquad d = p+q \, ,
 \label{eq:sym-break}
\end{align}
of the full conformal symmetry. 
In the previous formula $p$ is the dimension of the defect, while $q$ is its codimension.
The symmetry breaking \eqref{eq:sym-break} implies that correlation functions in defect CFT are not as constrained as in homogeneous CFT.
In particular, the two-point function \eqref{eq:two-pt-with-def} depends on two conformal cross-ratios $r$ and $w$, defined as in \cite{Lemos:2017vnx}
\begin{align}
 r + \frac{1}{r} 
 = \frac{|x_{12}^{\|}|^2 + |x_1^\bot|^2 + |x_2^\bot|^2}{|x_1^\bot| |x_2^\bot|} 
 \, , \qquad
 w + \frac{1}{w}
 = \frac{2 x_1^\bot \cdot x_2^\bot}{|x_1^\bot| |x_2^\bot|} 
 \, .
 \label{eq:cross-ratios}
\end{align}
Here we are considering a flat defect for simplicity, and we take $x^{\|}$ to be the directions parallel to the defect and $x^{\bot}$ the orthogonal directions.
For boundary or interface CFT, namely for codimension $q=1$, the second cross-ratio is constant $w = 1$ and the correlator only depends on $r$.
Since $w$ plays a central role in our story, from now on we assume that the defect is at least of codimension $q \ge 2$.
We have made some comments on boundary CFT in section \ref{subsec:bcft}.

The correlator $F(r,w)$ contains an infinite amount of CFT data which is extracted using the operator product expansion (OPE). 
In defect CFT there are two different expansions, the bulk and the defect OPE, each of them with their own CFT data
\begin{align}
\label{eq:cross-eq}
 F(r,w)
 = \left(\frac{r w}{(r-w) (r w-1)}\right)^{\frac{\Delta_{1}+\Delta_2}{2}} \sum_{\Om} c_\Om f_{\Delta,\ell}(r,w)
 = \sum_{\wh\Om} \mu_{\wh\Om} \wh f_{\wh\Delta,s}(r,w) \, .
\end{align}
Here $\{c_\Om, \Delta, \ell \}$ is the data extracted using the bulk expansion, while $\{\mu_{\wh\Om}, \wh\Delta, s\}$ is the one appearing in the defect expansion, and $f_{\Delta,\ell}$ and $\wh f_{\wh\Delta,s}$ are the conformal blocks, which have been studied in detail in the literature \cite{Billo:2016cpy,Lauria:2017wav,Lauria:2018klo,Isachenkov:2018pef,Liendo:2019jpu}.
The goal of the defect bootstrap program is to determine $F(r,w)$ from consistency of the bulk and defect expansions in \eqref{eq:cross-eq}.

This work provides several examples of how to solve the crossing equation analytically.
The main tool is the dispersion relation \eqref{eq:disp-rel}, that reconstructs the correlation function $F(r, w)$ as an integral of its discontinuity, and which we reproduce here for ease of reference:
\begin{align}
 F(r,w)
 = \int_0^r \frac{dw'}{2 \pi i}
 \frac{w (1 - w') (1 + w')}{ w' (w'-w) (1 - w w')}
 \Disc F(r,w') \, .
\end{align}
This non-perturbative formula applies to defect CFT with codimension $q>2$.
Subtleties related to codimension-two defects are discussed in section \ref{subsec:specialcase}, whereas subtleties related to convergence and subtractions are discussed in section \ref{subsec:subtractions}.
The dispersion relation \eqref{eq:disp-rel} is the defect CFT analog of \cite{Carmi:2019cub}, the main difference being that here $F(r,w)$ is reconstructed from a single rather than double discontinuity:
\begin{align}
 \Disc F(r,w') 
 = F(r, w' + i0) - F(r, w' - i0) \, .
 \label{eq:disc-def}
\end{align}
The dispersion relation \eqref{eq:disp-rel} is powerful because in many applications the discontinuity is simpler than the full correlator.
Indeed, the discontinuity can be computed using the bulk OPE in \eqref{eq:cross-eq}, and for each term in the expansion one finds \cite{Lemos:2017vnx,Gimenez-Grau:2021wiv,Barrat:2021yvp}
\begin{align}
\label{eq:disc-compute}
 \Disc \bigg[
 \left(\frac{r w}{(r-w) (r w-1)}\right)^{\frac{\Delta_{1}+\Delta_2}{2}} 
 f_{\Delta,\ell}(r,w) \bigg]
 \propto 
 \sin\left( \frac{\Delta-(\Delta_1+\Delta_2+\ell)}{2} \pi \right) \, .
\end{align}
As a result, for bulk operators near the multi-twist dimension $\Delta = \Delta_1 + \Delta_2 + \ell + 2n + \gamma$ the contribution to the discontinuity goes like a generally small anomalous dimension $\Disc \sim \gamma$. 
For CFTs with a small coupling $g$, typically the anomalous dimension goes like the coupling $\gamma \sim g$, so many of the contributions \eqref{eq:disc-compute} are suppressed by powers of the small coupling.
Even for non-perturbative CFT, the lightcone bootstrap \cite{Fitzpatrick:2012yx,Komargodski:2012ek} implies that $\gamma$ is small at large spins, so the contribution of large-spin operators is suppressed by powers of $1/\ell$.
These two examples justify our claim that $\Disc F$ is a simpler object than the full correlator.
In sections \ref{sec:monodromy} and \ref{sec:newresults} we show how this simplicity can be used to our advantage to bootstrap correlators in defect CFT with the dispersion relation \eqref{eq:disp-rel}.

\subsection{Derivation from Cauchy's integral formula}
\label{subsec:derivation-cauchy}

\begin{figure}[ht]
\centering
 \begin{tikzpicture}[baseline, scale=1.7]
  \draw[-stealth] (-1, 0) -- (3.5, 0);
  \draw[-stealth] (0, -1) -- (0, 3);
  \draw[snake it] (0, 0) -- (1, 0);
  \draw[snake it] (2, 0) -- (3.3, 0); 
  \draw (3.05, 3) -- (3.05, 2.55) -- (3.45, 2.55);
  \node at (1, -0.35) {$r$};
  \node at (2, -0.4) {$\frac{1}{r}$};
  \node at (3.3, 2.8) {$w'$};
  \node at (1.5, 1.5) {$w$};
  \node[circle, draw=black, fill=black, inner sep=0.75pt] at (1.5, 1.3) {};
  \draw[
    decoration={markings, mark=at position 0 with {\arrow{stealth}}},
    postaction={decorate}
  ] (1.5,1.4) circle (0.5);
  \draw[ 
    decoration={markings, mark=at position 0.5 with {\arrow{stealth}}},
    postaction={decorate}
  ]  (0, 0.15) -- (1, 0.15);
  \draw (1, 0.15) to[out=0,in=0] (1, -0.15);
  \draw[ 
    decoration={markings, mark=at position 0.5 with {\arrow{stealth}}},
    postaction={decorate}
  ]  (1, -0.15) -- (0, -0.15);
  \draw (0, -0.15) to[out=180, in=180] (0, 0.15);
  \draw[ 
    decoration={markings, mark=at position 0.5 with {\arrow{stealth}}},
    postaction={decorate}
  ]  (3.3, -0.15) -- (2, -0.15);
  \draw (2, -0.15) to[out=180, in=180] (2, 0.15);
  \draw[ 
    decoration={markings, mark=at position 0.5 with {\arrow{stealth}}},
    postaction={decorate}
  ]  (2, 0.15) -- (3.3, 0.15);
  \node at (1.0, 2.00) {$C_0$};
  \node at (0.5, 0.35) {$C_1$};
  \node at (2.5, 0.35) {$C_2$};
 \end{tikzpicture} 
 \caption{Summary of integration contours in the complex $w'$ plane.}
 \label{fig:complex-w}
\end{figure}
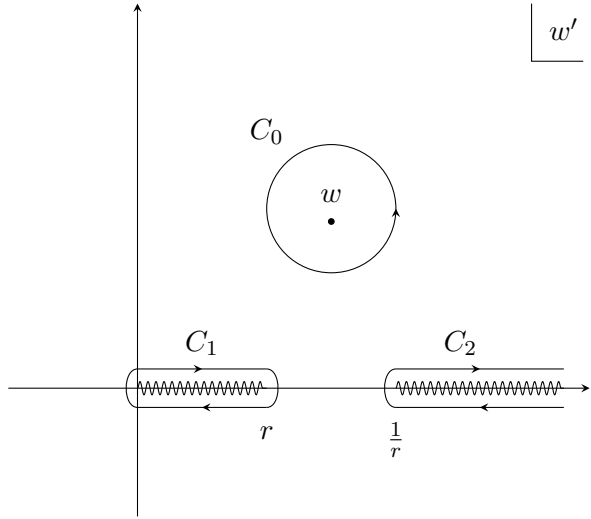

The simplest derivation of the dispersion relation follows from the analytic structure of the correlator $F(r,w)$ in the complex $w$ plane.
This has been discussed in \cite{Lemos:2017vnx}, where the authors conclude that for fixed $r$ there are two branch cuts $w \in (0,r)$ and $w \in (1/r, +\infty)$, see figure \ref{fig:complex-w}.
Starting with Cauchy's integral formula, and deforming the contour around the two cuts, we find
\begin{align}
 F(r,w)
 = \oint_{C_0} \frac{dw'}{2\pi i} \frac{F(r,w')}{w'-w}
 = \oint_{C_1+C_2} \frac{dw'}{2\pi i} \frac{F(r,w')}{w'-w} \, ,
 \label{eq:disp-rel-intermediate}
\end{align}
where the contours $C_i$ are defined in figure \ref{fig:complex-w}.
For now we are assuming the arcs at infinity can be dropped safely, but we shall reexamine this assumption in section \ref{subsec:subtractions}.

The right-hand side of \eqref{eq:disp-rel-intermediate} is the result we wanted to derive, although it can be massaged into a simpler form.
For this simplification we need $F(r,w') = F(r,1/w')$, which is true whenever the codimension is higher than two $q>2$, but also for parity-preserving codimension-two defects.
Then, the change of coordinates $w' \to 1/w'$ gives
\begin{align}
 \oint_{C_2} \frac{dw'}{2\pi i} \frac{F(r,w')}{w'-w}
 = - \oint_{C_1} \frac{dw'}{2\pi i} \frac{F(r,w')}{w'(1-w w')} \, .
 \label{eq:disp-rel-intermediate-2}
\end{align}
Combining the integration kernels \eqref{eq:disp-rel-intermediate} and \eqref{eq:disp-rel-intermediate-2}, we conclude
\begin{align}
 F(r,w)
 = \oint_{C_1} \frac{dw'}{2 \pi i}
 \frac{w (1 - w') (1 + w')}{ w' (1 - w w')}
 \frac{F(r,w')}{w'-w} \, .
 \label{eq:contour-disp-rel}
\end{align}
The final step is to rewrite the contour integral as a real integral of the discontinuity $\Disc F(r,w')$ in $w' \in (0,r)$, giving the dispersion relation \eqref{eq:disp-rel}.

Finally, let us note that in certain applications the contour integral \eqref{eq:contour-disp-rel} is more convenient than its real counterpart \eqref{eq:disp-rel}. 
This happens when, instead of a branch cut, the correlator $F(r,w')$ has a pole at $w' = r$. 
In this case the integral \eqref{eq:contour-disp-rel} can be computed using the residue theorem
\begin{align}
 F(r,w)
 = - \underset{w'=r}{\Res} \left[ 
 \frac{w (1 - w') (1 + w')}{ w' (1 - w w')}
 \frac{F(r,w')}{w'-w} \right] \, .
 \label{eq:pole-disp-rel}
\end{align}
This last formula will be useful in section \ref{sec:newresults}, in our study of  supersymmetric defects in $\Nm=4$ SYM.

\subsection{Derivation from Lorentzian inversion formula}
\label{subsec:derivation-lif}

Previous analytic studies \cite{Gimenez-Grau:2021wiv,Barrat:2021yvp} of two-point functions in defect CFT relied on a Lorentzian inversion formula \cite{Caron-Huot:2017vep,Lemos:2017vnx}.
In this section, we show that our dispersion relation captures the same information.
To prove this, we rederive our dispersion relation by resumming the Lorentzian inversion formula, similarly to how \cite{Carmi:2019cub} obtained their dispersion relation for CFT without defects.

Throughout this section we use results and notation from \cite{Lemos:2017vnx}, where the reader can find further details.
Our starting point is the decomposition of the correlator in terms of orthogonal functions:
\begin{equation}
F(r, w)=\sum_{s=0}^{\infty} \int_{p/2-i\infty}^{p/2+i\infty} \frac{d \widehat{\Delta}}{2 \pi \mathrm{i}} b(\widehat{\Delta}, s) \widehat{g}_{s}(w) \Psi_{\widehat{\Delta}}(r) \, .
\label{eq:defectexp}
\end{equation}
These orthogonal functions can be separated into a parallel factor
\begin{align}
 \Psi_{\widehat{\Delta}}(r) 
 & = \frac{1}{2} \left(
     \wh f_{\wh \Delta}(r)
   + \frac{K_{p-\wh\Delta}}{K_{\wh\Delta}} \wh f_{p-\wh\Delta}(r)
 \right) \, , \label{eq:defPsi} \\
\wh f_{\wh\Delta}(r)
 & = r^{\wh \Delta } 
    {}_2F_1\left(\wh \Delta, \frac{p}{2}, \wh \Delta - \frac{p}{2}+1,r^2\right) \, , \label{eq:def-block-rad} \\
 K_{\widehat{\Delta}}
 & = \frac{\Gamma(\widehat{\Delta})}{\Gamma \! \left(\widehat{\Delta}-\frac{p}{2}\right)} \, ,
\end{align}
and into an angular factor, which reads
\begin{align}
\widehat{g}_{s}(w) 
&= w^{-s}{ }_{2} F_{1}\left(-s, \frac{q}{2}-1,2-\frac{q}{2}-s, w^{2}\right) \,.
\label{eq:def-block-ang}
\end{align}
The function $b(\wh\Delta,s)$ encodes the defect CFT data in its poles and residues.
It follows from the expansion \eqref{eq:defectexp} and orthogonality of the basis functions, that $b(\wh\Delta,s)$ can be expressed as a double integral of $F(r,w)$.
The integration contour can be deformed such that only $\Disc F(r,w)$ contributes, leading to the Lorentzian inversion formula \cite{Lemos:2017vnx}:
\begin{align}
 b(\widehat{\Delta}, s)
 = & -\frac{2K_{\widehat{\Delta}}}{K_{p-\widehat{\Delta}}} 
   \int_{0}^{1} \frac{d r'}{r'} \int_{0}^{r'} \frac{d w'}{2\pi\mathrm{i} \, w'} 
   \notag \\
 & \times \left(\frac{1-w'^2}{w'} \right)^{q-2} \left(\frac{1- r'^2}{r'} \right)^{p} 
   \wh g_{2-s-q}(w') \Psi_{\widehat{\Delta}}(r') \Disc F(r', w')\,.
\label{eq:inversion}
\end{align}
Inserting the inversion formula \eqref{eq:inversion} in the orthogonal expansion \eqref{eq:defectexp} and swapping integrations and sums, we obtain an expression that resembles our dispersion relation
\begin{align}
 F(r,w)
 = \int_0^1 dr' \int_0^{r'} \frac{dw'}{2\pi i} 
 K_{\text{parallel}}(r,r') K_{\text{angular}}(w,w') \Disc F(r',w') \, .
\end{align}
The kernel is a product of parallel and angular terms, as follows from the factorized form of the basis functions in \eqref{eq:defectexp}.
The angular part is the easiest to compute, and gives
\begin{align}
 K_{\text{angular}}(w,w')
 & = \frac{-1}{w'} \left(\frac{1-w'^2}{w'}\right)^{q-2}
     \sum_{s=0}^\infty \wh g_s(w) \wh g_{2-s-q}(w') \notag \\
 & = \frac{w \left(1-w'^2\right)}{w' (w'-w) (1-w w')} \, . \label{eq:spin-sum}
\end{align}
This can be checked analytically comparing the two expansions around $w,w' = 0$  order by order, or numerically to very high precision by truncating the sum to a few thousand terms.

The calculation of the parallel kernel is somewhat more involved, but the final result consists only of a delta function:
\begin{align}
 K_{\text{parallel}}(r,r')
 = \frac{1}{r} \left(\frac{1- r'^2}{r'} \right)^{p} 
 \int_{p/2-i\infty}^{p/2+i\infty} \frac{d\widehat{\Delta}}{2 \pi i}
 \frac{2K_{\wh \Delta}}{K_{p-\wh \Delta}} \Psi_{\wh \Delta}(r) \Psi_{\wh \Delta}(r') 
 = \delta (r-r')\, .
 \label{eq:delta-int}
\end{align}
Multiplying the angular factor \eqref{eq:spin-sum} and the radial factor \eqref{eq:delta-int}, we obtain again the kernel in the dispersion relation \eqref{eq:disp-rel}.

Before concluding the section, let us outline how to obtain the radial kernel \eqref{eq:delta-int}.
We start expanding $\Psi_{\wh \Delta}(r)$ using \eqref{eq:defPsi}, so we get a sum of four terms of the form $f_{\wh\Delta_1}(r)f_{\wh\Delta_2}(r')$.
For each of these terms, we deform the integration contours to the left or right, according to the asymptotic behavior of blocks at large $\wh \Delta$:
\begin{align}
 \wh f_{\wh \Delta}(r) 
 \sim r^{\wh \Delta} (1-r^2)^{-p/2} \, .
 \label{eq:asymptotics-blocks}
\end{align}
For example, for the term $f_{\wh \Delta}(r) f_{\wh \Delta}(r')$, equation \eqref{eq:asymptotics-blocks} implies we must deform the contour to the right, while for $f_{p-\wh \Delta}(r) f_{p-\wh \Delta}(r')$ we must deform to the left.
For the crossed terms, we deform to the left or right depending on whether $r < r'$ or $r > r'$.
All in all, whenever $r \ne r'$ these four terms cancel.
When $r = r'$ there is a delta-function singularity, which can be obtained from the crossed terms expanded at large $\wh \Delta$, and then using the following integral:
\begin{align}
 \int_{p/2 - i \infty}^{p/2 + i\infty} d\a \, x^\a 
 = 2 \pi i \, \delta(x-1) \, .
\end{align}
The final result is \eqref{eq:delta-int}, where the delta function allows us to replace $r' = r$.

\subsection{The special case \texorpdfstring{$q=2$}{q=2}}
\label{subsec:specialcase}

The kinematics of codimension-two defects is somewhat special, as emphasized in \cite{Gimenez-Grau:2021wiv}.
The main peculiarity is the existence of parity-odd one-point tensor structures, which results in correlation functions without the symmetry $F(r,1/w) = F(r,w)$.
It is therefore not possible to combine the $C_1$ and $C_2$ contours in \eqref{eq:disp-rel-intermediate}, and the final dispersion relation is simply
\begin{align}
 F(r,w)
 = \int_0^r \frac{dw'}{2\pi i} \frac{\Disc F(r,w')}{w'-w} 
 + \int_{1/r}^{\infty} \frac{dw'}{2\pi i} \frac{\Disc F(r,w')}{w'-w} \, .
 \label{eq:disp-rel-q2}
\end{align}

Another special feature of codimension two is the existence of monodromy defects \cite{Billo:2013jda,Gaiotto:2013nva,Soderberg:2017oaa,Giombi:2021uae,Bianchi:2021snj,Gimenez-Grau:2021wiv}.
Monodromy defects are defined so that the external scalars pick a phase when going around defect:
\begin{align}
 \phi(r,\theta + 2\pi,\vec y) = e^{2\pi i v} \phi(r,\theta,\vec y) \, , \qquad
 \bar \phi(r,\theta + 2\pi,\vec y) = e^{-2\pi i v} \bar \phi(r,\theta,\vec y) \, .
 \label{eq:bc-monodromy}
\end{align}
Here $r,\theta$ are polar coordinates in the plane orthogonal to the defect, while $\vec y$ are parallel directions.
As a result of \eqref{eq:bc-monodromy}, the two-point function $\vvev{ \! \phi \bar \phi \! } \propto F(r,w)$ satisfies the boundary condition
\begin{align}
 F(r,e^{2\pi i} w) = e^{2\pi i v} F(r, w) \, .
\end{align}
Because of the monodromy, the correlator has an extra branch cut $w \in (-\infty, 0)$, and the contour manipulations of section \ref{subsec:derivation-cauchy} are invalid.
However, we can still apply the $q=2$ dispersion relation \eqref{eq:disp-rel-q2} to $w^{-v} F(r,w)$, which does not have the extra branch cut. 
As a result, the dispersion relation for monodromy defects is
\begin{align}
 F(r,w)
 = \int_0^r \frac{dw'}{2\pi i} 
   \left(\frac{w}{w'}\right)^v \frac{\Disc F(r,w')}{w'-w} 
 + \int_{1/r}^{\infty} \frac{dw'}{2\pi i}
   \left(\frac{w}{w'}\right)^v \frac{\Disc F(r,w')}{w'-w} \, ,
 \label{eq:disp-rel-mon}
\end{align}
where we remind the reader that $v$ parametrizes the monodromy of the external scalar $\phi(x)$ when it circles around the defect, see \eqref{eq:bc-monodromy}.
We have also derived \eqref{eq:disp-rel-mon} by resumming the Lorentzian inversion formula, as in section \ref{subsec:derivation-lif}.
However, it is crucial to use the inversion formula derived in \cite{Gimenez-Grau:2021wiv}, which takes into account the subtleties of codimension-two defects.

\subsection{Subtractions}
\label{subsec:subtractions}

So far, we have assumed that the correlator $F(r,w)$ decays fast enough in the complex-$w$ plane for the dispersion relations to be valid.
However, in realistic theories this assumption might not be true.
In this section we describe how to combine the dispersion relation with subtractions, in order for it to be applicable in the most general situations.

If we review the derivation in section \ref{subsec:derivation-cauchy}, we notice that equation \eqref{eq:disp-rel-intermediate} assumes that the arcs at infinity do not contribute.
However, in general our correlator goes as
\begin{align}
 |F(r,w)| < K |w|^{s_*} \quad
 \text{as} \quad |w| \to \infty \, ,
\end{align}
and only if $s_* < 0$ our dispersion relation \eqref{eq:disp-rel} is valid.
Let us relax this assumption, and instead imagine that $s_* \ge 0$.
Our goal is to find a subtracted correlator $\tilde F(r,w)$ with the same singularity structure as $F(r,w)$, but that decays fast enough at infinity
\begin{align}
 |\tilde F(r,w)| < K |w|^{-\veps} \quad
 \text{as} \quad |w| \to \infty \, , \quad \veps > 0 \, .
 \label{eq:subtracted-asympt}
\end{align}
If we find such $\tilde F(r,w)$, then the dispersion relation reconstructs it from its discontinuity.
In order to obtain $\tilde F(r,w)$, one subtracts from $F(r,w)$ a (possibly infinite) number of defect blocks $\wh f_{\wh\Delta,s}$ with $s \le s_*$.
More precisely, the subtracted correlator is
\begin{align}
 \tilde F(r,w)
 = F(r,w) 
 - \sum_{s=0}^{s_*} \sum_{i} \mu_{s,i} \wh f_{\wh\Delta_i,s}(r,w) \, ,
\end{align}
where the scaling dimension $\wh\Delta_i$ and coefficients $\mu_{s,i}$ are chosen to ensure \eqref{eq:subtracted-asympt}.
The defect conformal blocks are $\wh f_{\wh\Delta,s}(r,w) = \wh f_{\wh \Delta}(r) g_s(w)$, with the radial and angular part defined in \eqref{eq:def-block-rad} and \eqref{eq:def-block-ang} respectively.
As a result, they are polynomials in $w + w^{-1}$ of degree $s$, so the subtraction does not change the discontinuity
\begin{align}
 \Disc \tilde F(r,w) = \Disc F(r,w) \, .
\end{align}
All in all, we conclude that it is always possible to use a subtracted dispersion relation that reads
\begin{align}
 F(r,w)
 = \int_0^r \frac{dw'}{2 \pi i}
 \frac{w (1 - w') (1 + w')}{ w' (w'-w) (1 - w w')}
 \Disc F(r,w')
 + \sum_{s=0}^{s_*} \sum_i \mu_{s,i} \wh f_{\wh\Delta_i,s}(r,w) \, .
 \label{eq:disp-rel-subt}
\end{align}
In an actual bootstrap calculation, one often knows $\Disc F$, but extra work is needed to determine whether the dispersion relation needs subtractions, namely whether $s_* < 0$.
In case subtractions are needed, it is generally a non-trivial task to determine $\{\wh\Delta_i, \mu_i\}$.
In certain models, such as $\Nm=4$ SYM at strong coupling, it is possible to determine the subtractions as we show in section \ref{sec:newresults}.

\section{Monodromy defects}
\label{sec:monodromy}

In this section, we apply the dispersion relation to monodromy defects, see \cite{Gaiotto:2013nva,Soderberg:2017oaa,Giombi:2021uae,Bianchi:2021snj,Gimenez-Grau:2021wiv} for previous work.
We consider three examples, namely a trivial defect, a generalized free field, and the Wilson-Fisher fixed point.
In the three setups, we reproduce results previously obtained with the Lorentzian inversion formula \cite{Gimenez-Grau:2021wiv}, and we demonstrate how the dispersion relation significantly simplifies the calculations.
This section serves as a sanity check for our formalism, as well as to illustrate how to employ it in simple setups.
In principle our methods could also be applied to other setups, such as the line defects of \cite{Cuomo:2021kfm,Cuomo:2022xgw} in the $\veps$--expansion, but we leave this to future work.

\subsection{Trivial defect}
\label{subsec:bulkid}

The simplest correlator to consider is that of a trivial defect.
A defect is called trivial if it does not interact with the bulk, meaning that all bulk operators except the identity have vanishing one-point functions $\vvev{ \! \Om \! } = 0$.%
\footnote{Alternatively, we can think of it as a monodromy defect with a trivial monodromy $v=0$.}
As a result, the bulk-channel expansion contains only the identity operator, and the full two-point function reads
\begin{align}
 F_{\text{trivial}}(r,w) 
 = \left(\frac{r w}{(w-r) (1-r w)}\right)^{\Delta_{\phi}} \, .
 \label{eq:blkid}
\end{align}
Our goal is to show that the dispersion relation reconstructs this correlator starting from its discontinuity.
The discontinuity follows from the definition \eqref{eq:disc-def}
\begin{align}
 \frac{1}{2\pi i} \, \underset{0<w'<r}{\Disc} F_{\text{trivial}}(r,w')
 = - \frac{ \sin (\pi  \Delta_{\phi})}{\pi} 
 \left(\frac{r w'}{(r-w') (1-r w')}\right)^{\Delta_{\phi}} \,.
\end{align}
Thus, according to the dispersion relation \eqref{eq:disp-rel}, the following identity should hold:
\begin{align}
 F_{\text{trivial}}(r,w)
 \stackrel{?}{=}
 - \frac{ \sin (\pi  \Delta_{\phi})}{\pi}
 \int_{0}^r dw' \frac{w (1 - w') (1 + w')}{ w' (w'-w) (1 - w w')}
 \left(\frac{r w'}{(r-w') (1-r w')}\right)^{\Delta_{\phi}} \, .
 \label{eq:dispidentity}
\end{align}
We have not been able to compute this integral analytically. 
Instead, we have changed variables $w' \to w' r$, and we then expanded the integrand for $r \ll 1$. At each order in $r$ the integrals in $w'$ are elementary, giving for the first few terms
\begin{align}
 F_{\text{trivial}}(r,w)
 = r^{\Dp} \left(
     1
   + \Dp r \left(w+w^{-1}\right)
   + \frac{1}{2} \Dp r^2 \left((\Dp+1)\left(w+w^{-1}\right)^2-2\right)
   + O(r^3)
 \right) .
\end{align}
Continuing this process to high order in $r$, we observe that the dispersion integral \eqref{eq:dispidentity} reproduces the Taylor series of \eqref{eq:blkid}.

\subsection{Monodromy defect for generalized free theory}
\label{subsec:free-monodromy}

The next example we consider is the two-point function of generalized free fields (GFF) in the presence of a monodromy defect
\begin{align}
 \vvev{ \phi(x_1) \bar \phi(x_2) }
 = \frac{F^{\text{GFF}}_v(r,w)}{|x_1^\bot|^\Dp |x_2^\bot|^\Dp} \, .
\end{align}
The external field is a complex scalar $\phi$ that picks a monodromy $\phi \to e^{2\pi i v} \phi$ when circling around the defect.
Because $\phi$ is a generalized free field, its operator product expansion contains the identity followed by operators with exact multi-twist dimension
\begin{align}
 \phi \times \bar \phi \sim \mathds{1} + \sum_{\ell,n} \Om_{\ell,n} \, , \qquad
 \Delta_{\ell,n} = 2 \Dp + \ell + 2n \, .
\end{align}
As discussed around equation \eqref{eq:disc-compute}, the discontinuity kills operators with multi-twist dimension, which means only the bulk identity contributes.
Because this is a monodromy defect, see section \ref{subsec:specialcase}, we need to compute the discontinuity across the two possible branch cuts, namely
\begin{align}
 - \, \underset{0<w'<r}{\Disc} F_{v}^{\text{GFF}}(r,w')
 = \underset{1/r<w'}{\Disc} F_{v}^{\text{GFF}}(r,w')
 = 2\pi i \, \frac{ \sin (\pi  \Delta_{\phi})}{\pi} 
 \left(\frac{r w'}{(r-w') (1-r w')}\right)^{\Delta_{\phi}} .
\end{align}
Inserting the two discontinuities in the dispersion relation for monodromy defects \eqref{eq:disp-rel-mon}, gives an integral formula for the correlator:
\begin{align}
 F^{\text{GFF}}_{v}(r,w)
 = r^{\Dp} w^v \frac{\sin (\pi  \Dp)}{\pi} \Bigg( 
   & - \int_0^r dw' 
       \frac{w'^{\Dp-v}}{(w'-w) (r-w')^{\Dp} (1-r w')^{\Dp}} \notag \\*
   & + \int_{1/r}^\infty dw' 
       \frac{w'^{\Dp-v}}{(w'-w) (w'-r)^{\Dp} (r w'-1)^{\Dp}} 
 \Bigg) \, .
\end{align}
Once again, it is a complicated task to compute these integrals exactly.
Instead, we expand for $r \ll 1$ and integrate term by term, producing the following double sum
\begin{align}
 F^{\text{GFF}}_{v}(r,w)
 = \frac{r^{\Dp}}{\Gamma(\Dp)}
 \sum_{n,m=0}^\infty \frac{(\Dp)_n}{n!} & \bigg(
   \frac{\Gamma (m+n+v+\Dp)}{\Gamma (m+n+v+1)} w^{m+v} r^{m+2 n+v} \notag \\*
&+ \frac{\Gamma (m+n-v+\Dp+1)}{\Gamma (m+n-v+2)} w^{-m+v-1} r^{m+2 n-v+1}
 \bigg) \, .
\end{align}
Each term in the double sum has the form of an Appell $F_1$ series, leading to the final result\footnote{We thank Y.~Linke for bringing this closed formula to our attention. We also acknowledge interesting discussions with S.~Dowker, who rederived this result with alternative methods \cite{Dowker:2022mex}.}
\begin{align}
 F^{\text{GFF}}_{v}(r,w)
 = \, & \, \frac{r^{\Dp}}{\Gamma (\Dp)} \Bigg(
   \frac{\Gamma (v+\Dp)}{\Gamma (v+1)}
   (r w)^v
   F_1\left(v+\Dp;1,\Dp;v+1;r w,r^2\right) \notag \\*
&+ \frac{\Gamma (\Dp+1-v)}{\Gamma (2-v)} 
   \left(\frac{r}{w}\right)^{1-v}
   F_1\left(\Dp+1-v;1,\Dp;2-v;\frac{r}{w},r^2\right)
 \Bigg) \, .
 \label{eq:gff-monodromy-corr}
\end{align}
This represents an improvement to the results presented in \cite{Gimenez-Grau:2021wiv}, where this formula could only be obtained in certain limits.
In particular, one can expand equation \eqref{eq:gff-monodromy-corr} in bulk-channel blocks to extract the bulk CFT data of GFF monodromy defects, which is currently unknown.

\subsection{Monodromy defect in the Wilson-Fisher fixed point}
\label{subsec:wf-monodromy}

Another interesting example is the monodromy defect in the Wilson-Fisher fixed point.
This setup has been studied using diagrammatic perturbation theory \cite{Gaiotto:2013nva,Soderberg:2017oaa,Giombi:2021uae} and analytic bootstrap \cite{Liendo:2019jpu,Gimenez-Grau:2021wiv}.
The analytic bootstrap is particularly simple, because only the identity and the operator $\phi^2$ contribute to the discontinuity at leading order in $\veps$ \cite{Gimenez-Grau:2021wiv}:
\begin{align}
 \Disc F(r,w) \big|_{O(\veps)}
 = \Disc \left(\frac{r w}{(w-r) (1-r w)}\right)^{\Delta_{\phi}} \left[
    1 + f_{\Delta_{\phi^2}, 0}(r, w)
 \right] \, .
\end{align}
Therefore, the full correlator at order $O(\veps)$ can be reconstructed using the dispersion relation.
The correlator is naturally split as a sum of two terms. 
The first term is obtained from the identity, and it has been considered in section \ref{subsec:free-monodromy}.
The second term, which we call $F_{\text{WF}}$, is the focus of the current example.
This term is obtained from the discontinuity of $f_{\Delta_{\phi^2,0}}$, that up to an unimportant overall normalization reads \cite{Gimenez-Grau:2021wiv}
\begin{align}
 - \underset{0<w<r}{\Disc} F_{\text{WF}}(r,w)
 = \underset{1/r<w}{\Disc} F_{\text{WF}}(r,w)
 = 2 \pi i \, \frac{r \log r}{1-r^2} \, .
\end{align}
The steps to reconstruct $F_{\text{WF}}$ are the same as in the previous examples.
After inserting the discontinuity in the dispersion relation \eqref{eq:disp-rel-mon}, we expand the integrand for $r \ll 1$ and integrate term by term.
The resulting infinite sum can be computed in closed form, giving:
\begin{align}
 F_{\text{WF}}(r,w)
&= \frac{r \log r}{1-r^2} \Bigg[ 
 - \int_0^r dw' 
   \left(\frac{w}{w'}\right)^v \frac{1}{w'-w} 
 + \int_{1/r}^{\infty} dw'
   \left(\frac{w}{w'}\right)^v \frac{1}{w'-w} 
 \Bigg] \notag \\[0.5em]
&= \frac{r \log r}{1-r^2} \sum_{m=0}^\infty \Bigg[ 
     \frac{(r/w)^{m+1-v}}{m+1-v}
   + \frac{(r w)^{m+v}}{m+v}
 \Bigg] \notag \\[0.5em]
&= \frac{r \log r}{1-r^2} \bigg[ 
    \left( \frac rw \right)^{1-v} \Phi\left(\frac{r}{w},1,1-v\right)
   + (r w)^v \Phi(r w,1,v)
 \bigg] \, .
\end{align}
The result, which is expressed with the Lerch zeta function $\Phi(x,s,v) = \sum_{n=0}^\infty x^n  (n+v)^{-s}$, is once again in perfect agreement with the literature \cite{Gimenez-Grau:2021wiv}.

\section{Wilson line defect in \texorpdfstring{$\Nm = 4$}{N=4} SYM}
\label{sec:newresults}

In this section we consider a supersymmetric Wilson line in planar $\Nm = 4$ SYM in the presence of single-trace half-BPS operators $\Om_J$.
In \cite{Barrat:2021yvp}, we proposed a method to bootstrap the two-point function $\vvev{\Om_J \Om_J}$ at leading non-trivial order in the strong-coupling regime.
With the help of the dispersion relation, we wish now to extend these results to the case of unequal external operators $\vvev{\Om_{J_1} \Om_{J_2} }$. 
After a brief summary of the setup, we proceed to bootstrap the simplest correlator $\vvev{ \Om_2 \Om_3 }$.
The method is completely general, and can be applied to more correlators, as we do in appendix \ref{sec:appendixA}.
We conclude this section with a conjecture on the form of the highest $R$-symmetry channel for arbitrary $J_1 \ne J_2$, which generalizes our previous conjecture \cite{Barrat:2021yvp}.

\subsection{Setup}
\label{subsec:setup}

In this section we gather the essential details about the setup. 
We follow the same conventions as in our previous work \cite{Barrat:2021yvp}, where the reader can find a more complete description.

The most important operators we consider are single-trace half-BPS operators, defined as
\begin{equation}
\Om_J (x,u) := (2\pi)^J \frac{2^{J/2}}{\sqrt{J \lambda^J}} \text{tr}\, \left( u \cdot \phi(x) \right)^J\,.
\end{equation}
Here $u$ is a six-dimensional polarization vector satisfying $u^2 = 0$ to ensure that $\Om_{J}(x,u)$ transforms in the symmetric-traceless representation of the $R$-symmetry group. These operators are protected and preserve half of the supercharges of the bulk theory. 
We shall study the half-BPS operators in the presence of a supersymmetric Wilson line defect, defined as
\begin{equation}
\Wl := \frac{1}{N}\, \text{tr}\, \Pm\, \exp \int_{-\infty}^{\infty} d\tau\, \left( i \dot{x}^\mu A_\mu + |\dot{x}|\, \theta \cdot \phi \right)\, .
\end{equation}
The six-component polarization vector $\theta$ satisfies $\theta^2 = 1$ and determines which scalar fields couple to the line.

In this setup, the two-point function of single-trace operators $\Om_{J_1}$ and $\Om_{J_2}$ takes the form
\begin{align}
 \vvev{ \Om_{J_1}(x_1, u_1) \Om_{J_2}(x_2, u_2) }
 = \frac{(u_1 \cdot \theta)^{J_1} (u_2 \cdot \theta)^{J_2}}{|x_1^\bot|^{J_1} |x_2^\bot|^{J_2}}
   \Fm(z,\zb,\sigma) \, ,
\end{align}
where the double-bracket notation indicates the correlator is taken in the presence of the supersymmetric Wilson line. 
The reduced correlator $\Fm(z,\zb,\sigma)$ depends on two spacetime cross-ratios $z, \zb$ and one $R$-symmetry variable $\sigma$.
To agree with the conventions of \cite{Barrat:2021yvp}, we use the spacetime cross-ratios $z$ and $\zb$, which are related to our previous definition \eqref{eq:cross-ratios} as
\begin{equation}
z = r w\,, \qquad \zb = \frac{r}{w}\,.
\end{equation}
In order to decompose the correlator into $R$-symmetry channels, it is natural to use the variable $\sigma$ defined as
\begin{equation}
\sigma = -\frac{(1-\alpha)^2}{2 \alpha} = \frac{u_1 \cdot u_2}{(u_1 \cdot \theta)(u_2 \cdot \theta)}\,.
\label{eq:invariant}
\end{equation}
In terms of $\sigma$ the correlator $\Fm(z,\zb,\sigma)$ is a polynomial of degree $\min(J_1,J_2)$, with each power corresponding to a different $R$-symmetry channel.
These $R$-symmetry channels are not independent, but instead they are related to each other by the superconformal Ward identities \cite{Liendo:2016ymz}:
\begin{equation}
\left. \left( \partial_z + \frac{1}{2} \partial_\alpha \right) 
\Fm(z,\zb,\sigma) \right|_{z = \alpha} =
\left. \left( \partial_{\zb} + \frac{1}{2} \partial_\alpha \right) 
\Fm(z,\zb,\sigma) \right|_{\zb = \alpha} = 0\, .
\label{eq:WI}
\end{equation}
These Ward identities are expressed naturally in terms of $\alpha$, whose relation to $\sigma$ appears in \eqref{eq:invariant}.

Let us also review the expansion of the correlator in conformal blocks. 
As usual, there are two channels one can consider. 
The first one is the defect channel
\begin{align}
\label{eq:defect-block}
\Fm(z,\zb,\sigma)
 = \sum_{\wh\Om} b_{J_1\wh\Om} b_{J_2\wh\Om} \, 
   \wh h_{\wh K}(\sigma) \wh f_{\Dh,s}(z,\zb) \, ,
\end{align}
where the spacetime and $R$-symmetry blocks are the same as for equal external operators:
\begin{align}
\fh_{\Dh,s}(z, \zb)
 = & z^{\frac{\Dh-s}{2}} \zb^{\frac{\Dh+s}{2}} 
 \, _2F_1\left(\frac{1}{2},-s;\frac{1}{2}-s;\frac{z}{\zb}\right) 
 \, _2F_1\left(\frac{1}{2},\Dh;\Dh+\frac{1}{2};z \zb\right) \, , \\
\wh h_{\wh K}(\sigma)
 = & \sigma^{\wh K} 
 \, _2F_1\left(-\wh K-1,-\wh K;-2 (\wh K+1); \frac{2}{\sigma}\right) \, .
\end{align}
The expansion coefficients $b_{J\wh\Om}$ can be obtained from the normalization of the bulk-defect correlator $\vvev{ \! \Om_J \wh \Om \! }$, with our precise conventions spelled out in \cite{Barrat:2021yvp}.
On the other hand, the bulk-channel expansion reads
\begin{align}
 & \Fm(z,\zb,\sigma)
 = \left( \frac{z \zb\, \sigma }{(1-z)(1-\zb)}\right)^{\frac{J_1 + J_2}{2}}
   \sum_{\Om} \lambda_{J_1J_2\Om} a_\Om \,
   h^{J_{12}}_K(\sigma) f^{J_{12}}_{\Delta,\ell}(z,\zb) \, ,
\label{eq:blkexp}
\end{align}
with the $R$-symmetry blocks 
\begin{align}
 h^{J_{12}}_K(\sigma)
 = \sigma^{-K/2} 
   \, _2F_1\left(-\frac{K + J_{12}}{2},-\frac{K - J_{12}}{2};-K-1; \frac{\sigma}{2} \right) \, ,
 \label{eq:R-sym-block}
\end{align}
now adapted for unequal external charges $J_1$ and $J_2$, and with $J_{12} := J_1 - J_2$. 
The spacetime blocks $f^{J_{12}}_{\Delta,\ell}(z,\zb)$ do not have a known closed formula in terms of elementary functions, but useful series expansions can be found for example in \cite{Billo:2016cpy,Lauria:2017wav,Isachenkov:2017qgn}.
The coefficients in the bulk expansion are a product of a three-point coefficient $\lambda_{IJ\Om}$, and a one-point function in the presence of the defect $a_\Om$. 

Note that the bulk and defect expansions are expressed in terms of non-supersymmetric conformal blocks, or in other words, the sums in \eqref{eq:defect-block} and \eqref{eq:blkexp} run over superprimary and superdescendant operators.
In principle, the contribution of superdescendants can be related to the superprimaries, so the bulk and defect expansions could be reformulated in terms of superconformal blocks.
For example, all the superconformal blocks are known for the correlator $\vvev{ \Om_2 \Om_2 }$, see the appendix of \cite{Barrat:2020vch} or the notebook attached to \cite{Barrat:2021yvp}.
However, for the purposes of this work we only need to know one type of superblock.
This is the superblock that corresponds to the exchange of a half-BPS operator $\Om_K$ in the OPE of $\Om_{J_1}$ and $\Om_{J_2}$:
\begin{align}
  \Gm^{J_{12}}_{\Om_K}
& = h^{J_{12}}_K(\sigma) f^{J_{12}}_{K,0}(z,\zb)
  + \frac{(J_{12}^2 - K^2)(J_{12}^2 - (K+2)^2)}{128K (K+1)^2(K+3)} 
   h^{J_{12}}_{K-2}(\sigma) f^{J_{12}}_{K+2,2}(z,\zb) \notag \\[0.2em]
& \quad
  + \frac{(J_{12}^2 - K^2)^2 [(J_{12}^2 - 4)^2-2K^2 (J_{12}^2+4)+K^4]}{16384 (K-2)(K-1)^2K^2(K+1)(K+2)(K+3)} h^{J_{12}}_{K-4}(\sigma) f^{J_{12}}_{K+4,0}(z,\zb) \,.
 \label{eq:stblocks}
\end{align}
The relative coefficients can be fixed demanding that the superconformal block satisfies the Ward identities \eqref{eq:WI}, as was done originally in \cite{Liendo:2016ymz}.
Note that setting $J_1 = J_2$ we recover the superblocks for equal operators given in \cite{Barrat:2021yvp}.

After these preliminaries, we are ready to summarize the steps to bootstrap the correlators.
We work in the planar limit $N \to \infty$, where we keep the 't Hooft coupling fixed but large $\lambda \gg 1$.
From the holographic description, it follows that the first three orders at strong coupling can be expressed in the following way:
\begin{align}
 \Fm(z,\zb,\sigma)
 = \delta_{J_1,J_2} \left( \frac{z \zb\, \sigma }{(1-z)(1-\zb)}\right)^{\frac{J_1 + J_2}{2}}
 + \frac{\sqrt{J_1 J_2}}{2^{\frac{J_1+J_2+4}{2}}} \frac{\lambda}{N^2}
 + \!\!\! \sum_{n=0}^{\min(J_1,J_2)} \!\!\! \sigma^{\min(J_1,J_2)-n} F_n(z,\zb) \, .
\end{align}
The first term is of order $O(1)$, and corresponds to a disconnected Witten diagram.
The second term is of order $O(\frac{\lambda}{N^2})$, and corresponds to a Witten diagram that factorizes in a product of one-point functions.
The third term is of order $O(\frac{\sqrt{\lambda}}{N^2})$, and captures the leading non-trivial correction to the correlator.
We expand the non-trivial correction in powers of $\sigma$, such that each function $F_n$ corresponds to one $R$-symmetry channel.
The functions $F_n$ are the ones we wish to bootstrap.

The main tool needed for the bootstrap is the dispersion relation \eqref{eq:disp-rel}, that reconstructs the full correlator starting from its discontinuity.
At the order we are interested in, the discontinuity can be computed by noting that only single-trace half-BPS operators contribute. 
Indeed, all other operators in the bulk OPE have double-trace dimensions, with zero anomalous dimensions at this order.
As a result of equation \eqref{eq:disc-compute}, these double-trace operators are killed by the discontinuity.
Summarizing, the discontinuity at order $O(\frac{\sqrt{\lambda}}{N^2})$ is obtained as a sum over single-trace half-BPS contributions:
\begin{equation}
\Disc \Fm (z, \zb, \sigma) 
= \Disc \left( \frac{z \zb\, \sigma }{(1-z)(1-\zb)}\right)^{\frac{J_1 + J_2}{2}} \,
   \sum_{j=\frac{1}{2}|J_{12}|}^{\frac{1}{2}(J_1 + J_2 -2)} \!\! \lambda_{J_1,J_2,2j} \, a_{2j} \,
   \Gm^{J_{12}}_{\Om_{2j}} (z, \zb, \sigma)\, .
\label{eq:discblocks}
\end{equation}
The ranges of the sum follow from $R$-symmetry conservation. 
Although the sum should also include the operator $\Om_{J_1+J_2}$, it has exact double-trace dimension and is hence also killed by the discontinuity. 
Because we choose to work with $SU(N)$ gauge group, the operator $\Om_1 \propto \tr \phi$ is identically zero, and it should also be removed from the sum whenever it appears.
However, if one works with $U(N)$ then the operator $\Om_1$ should be kept.

We are now in a good position to derive the correlator $\Fm(z,\zb,\sigma)$, since the OPE coefficients $\lambda_{J_1 J_2 J}$ and $a_J$ can be computed using localization, and the superconformal blocks are known.
All is left to do is to use the dispersion relation \eqref{eq:disp-rel} to reconstruct $\tilde \Fm(z,\zb,\sigma)$, which is a subtracted version of the full correlator, see the discussion in section \ref{subsec:subtractions}.
Adding low-spin contributions to the subtracted correlator and demanding consistency with the superconformal Ward identities we will unambiguosly fix the full result.

\subsection{Case \texorpdfstring{$(J_1, J_2) = (2,3)$}{(J1,J2)=(2,3)}}
\label{subsec:case23}

For concreteness, let us focus on the specific example $(J_1, J_2) = (2,3)$.  
The result at the first non-trivial order can be decomposed into three $R$-symmetry channels
\begin{equation}
\Fm(z,\zb,\sigma) \Big|_{O(\frac{\sqrt{\lambda}}{N^2})} 
= \sigma^2 F_0 (z,\zb)
+ \sigma F_1(z,\zb)
+ F_2(z,\zb) \, .
\end{equation}
In this case, only one half-BPS block contributes to the discontinuity \eqref{eq:discblocks}, namely
\begin{equation}
\Disc \Fm (z, \zb, \sigma) = \Disc \left( \frac{z \zb\, \sigma }{(1-z)(1-\zb)}\right)^{\frac{5}{2}}
\lambda_{233} a_3 \, \Gm^{J_{12}=-1}_{\Om_3} (z, \zb, \sigma)\,,
\label{eq:disc-23}
\end{equation}
where the coefficient $\lambda_{233} a_3$ can be fixed using supersymmetric localization (see equations (2.23) and (2.25) in \cite{Barrat:2021yvp}):
\begin{equation}
\lambda_{233} a_3 = \frac{3\sqrt{3}}{4} \frac{\sqrt{\lambda}}{N^2} + \ldots \,.
\end{equation}
Here and throuought the paper, whenever we discuss CFT data we use $\ldots$ to represent corrections in both $1/N^2$ and $1/\sqrt{\lambda}$.
In order to proceed, it is necessary to note that the right-hand side of \eqref{eq:disc-23} does not have a branch cut, but instead it has a pole in $(1-\zb)$.
As discussed around equation \eqref{eq:pole-disp-rel}, all we need to reconstruct the correlator is the residue at this pole:
\begin{equation}
\left( \frac{z \zb\, \sigma}{(1-z)(1-\zb)}\right)^{\frac{5}{2}} 
\Gm^{-1}_{\Om_3} \Big|_{\text{pole}}
= \frac{1}{1-\zb} \, \frac{2 z}{(1+\sqrt{z})^3  (1-\sqrt{z})} \left( 2 \sigma - \frac{1 + 4 \sqrt{z} + z}{3 (1+\sqrt{z})^2} \sigma^2 \right) \,.
\label{eq:pole-term}
\end{equation}
In order to obtain this result, we used the formula for the superblock \eqref{eq:stblocks}, as well as the series expansion for the spacetime block in \cite{Isachenkov:2018pef,Liendo:2019jpu} (see also appendix A of \cite{Barrat:2021yvp} for further details).
The pole \eqref{eq:pole-term} can be inserted into equation \eqref{eq:pole-disp-rel}, and picking the apropriate residue gives:
\begin{align}
\tilde{F}_0(z,\zb) & = - \frac{\sqrt{\lambda}}{N^2} \frac{\sqrt{3} \, z \zb \, \big( 1+ 4 \sqrt{z \zb} + z \zb \big)}{2(1-z)(1-\zb) (1+\sqrt{z \zb})^4}\,, \label{eq:prelres-0} \\
\tilde{F}_1(z,\zb) & = \frac{\sqrt{\lambda}}{N^2}\frac{ 3\sqrt{3} \, z \zb}{(1-z)(1-\zb)(1+\sqrt{z\zb})^2}\,, \label{eq:prelres-1} \\
\tilde{F}_2(z,\zb) & = 0\,.
\label{eq:prelres-2}
\end{align}
Here we would like to remark the relative simplicity with which \eqref{eq:prelres-0}-\eqref{eq:prelres-2} has been obtained.
To reproduce this result with the Lorentzian inversion formula, one should first extract the CFT data of the defect operators twist by twist performing an integral of the discontinuity times a kernel.
Plugging this CFT data in the defect-channel expansion, one would then resum a double power series in $z$ and $\zb$, see \cite{Barrat:2021yvp} for more details.
Instead, all we had to do now is to multiply \eqref{eq:pole-term} with the kernel in \eqref{eq:pole-disp-rel} and pick the residue.

Note that we put a tilde on the channels $\tilde F_n$ to anticipate that they are not the final result, but instead that they should be understood as a subtracted correlator.
The reason they cannot be the full result is that they do not satisfy the superconformal Ward identities \eqref{eq:WI}.
As discussed in section \ref{subsec:subtractions}, to recover the full correlator from the subtracted part, one has to add an a priori infinite number of defect blocks for $s \le s_*$, see \eqref{eq:disp-rel-subt}.
It was observed in \cite{Barrat:2021yvp} that in the present context the three channels have different value of $s_*$.
In particular, the highest $R$-symmetry channel $\tilde F_0$ converges (i.e. $s_* < 0$), while we have to add $s=0$ blocks to $\tilde F_1$ and $s=0,1$ blocks to $\tilde F_2$.
Since we are working in perturbation theory, we expect that conformal blocks and their derivatives can appear, and assuming that only integer-dimension operators contribute we get the ansatz:
\begin{align}
\begin{split}
F_0 (z,\zb) &= \tilde{F}_0 (z,\zb) \,, \\
F_1 (z,\zb) &= \tilde{F}_1 (z,\zb) + \sum_{n=0}^\infty \left( k_n \hat f_{n, 0}(z, \zb) + p_n \partial_{\Dh} \hat f_{n, 0}(z, \zb) \right)\, , \\
F_2 (z,\zb) &= \tilde{F}_2 (z,\zb) + \sum_{s=0,1} \sum_{n=0}^\infty \left( q_{n,s} \hat f_{n+s, s}(z, \zb) + r_{n,s} \partial_{\Dh} \hat f_{n+s, s}(z, \zb) \right)\,.
\end{split}
\label{eq:amb22}
\end{align}
At this stage the coefficients $k_n$, $p_n$, $q_{n,s}$ and $r_{n,s}$ are unknown, but we fix them imposing that the correlator is supersymmetric, namely that it satisfies the superconformal Ward identities \eqref{eq:WI}. 
For the channel $F_1$, the Ward identities fix all but the coefficient $k_1$:
\begin{align}
  F_1 (z,\zb) 
& = \tilde{F}_1 (z,\zb) 
  + \left( k_1 - \frac{3\sqrt{3}}{4} \frac{\sqrt{\lambda}}{N^2} \right) \tanh^{-1} \sqrt{z\zb} 
  + \frac{\sqrt{3} \sqrt{\lambda }}{4 N^2}
    \frac{\sqrt{z \zb} \left(3 z \zb+2 \sqrt{z\zb}+3\right)}{\left(1 + \sqrt{z\zb}\right)^4}
    \, .
\label{eq:F1tild}
\end{align}
Here one should note that the presence of $\tanh^{-1}\sqrt{z\zb}$ is at tension with our assumption that the bulk operators do not have anomalous dimensions at the order we are working.
The reason is that in its expansion near $z \to 1$, there are logarithmic terms $\log (1-\zb)$, which would give rise to anomalous dimensions for bulk operators.
Thus, we must make sure that this term vanishes by setting
\begin{equation}
k_1 = \frac{3\sqrt{3}}{4} \frac{\sqrt{\lambda}}{N^2}\,.
\label{eq:k1-fix}
\end{equation}
Another way to fix $k_1$ is to note that the correlator must be symmetric under $z,\zb \to \frac1z,\frac1\zb$, as follows from the definition of the cross-ratios \eqref{eq:cross-ratios}.
The only term in \eqref{eq:F1tild} not symmetric under this transformation is $\tanh^{-1}\sqrt{z\zb}$, so we conclude that \eqref{eq:k1-fix} should hold.

For the channel $F_2$, the Ward identities fix all the coefficients except the constant $q_{0,0}$:
\begin{align}
 F_2 (z,\zb) 
 = \frac{\sqrt{3} \sqrt{\lambda }}{4 N^2} 
   \frac{\sqrt{z \zb}}{(1 + \sqrt{z \zb})^4} \Big[ 
     z + \zb
     - 3 z \zb - 4 \sqrt{z \zb} - 3
   \Big]
 + q_{0,0} \,.
\end{align}
However, the constant $q_{0,0}$ corresponds to the exchange of the identity operator in the defect channel, and this coefficient can be fixed using localization:
\begin{equation}
 a_2 a_3 
 = \frac{\sqrt{3}}{16} \frac{\lambda}{N^2}
 - \frac{11 \sqrt{3}}{32} \frac{\sqrt{\lambda}}{N^2}
 + \ldots
 \quad \Rightarrow \quad
 q_{0,0} = - \frac{11 \sqrt{3}}{32} \frac{\sqrt{\lambda}}{N^2}\,.
\end{equation}

Putting everything together, we find the final result for the three $R$-symmetry channels:
\begin{align}
F_0(z,\zb) & = - \frac{\sqrt{3} \sqrt{\lambda}}{2 N^2} \frac{z \zb (1+ 4 \sqrt{z\zb} + z\zb)}{(1-z)(1-\zb)(1+\sqrt{z\zb})^4}\,, \\
F_1(z,\zb) 
& = \frac{\sqrt{3} \sqrt{\lambda}}{N^2} \frac{ \sqrt{z \zb}}{(1+\sqrt{z\zb})^2} \Bigg[
    \frac{ 3 \, \sqrt{z \zb}}{(1-z)(1-\zb)}
+ \frac{3 z \zb+2 \sqrt{z\zb}+3}
       {4 \left(1 + \sqrt{z\zb}\right)^2} \Bigg] \,, \\
F_2(z,\zb) 
& = - \frac{11 \sqrt{3}}{32} \frac{\sqrt{\lambda}}{N^2} 
  + \frac{\sqrt{3} \sqrt{\lambda }}{4 N^2} 
   \frac{\sqrt{z \zb}}{(1 + \sqrt{z \zb})^4} \Big[ 
     z + \zb
     - 3 z \zb - 4 \sqrt{z \zb} - 3
   \Big]\, .
\end{align}
It is interesting that there are no logarithmic terms in this result, as opposed to the case $\vvev{\Om_J \Om_J}$ in \cite{Barrat:2021yvp}. 
We believe this is a general feature of correlators with $J_{12}$ odd, while correlators with $J_{12}$ even always contain logarithms. 
In appendix \ref{sec:appendixA} we provide two additional examples for $(J_1, J_2) = (2,4)$ and $(J_1, J_2) = (3,4)$, which were obtained using the same method as here.

There are a few checks that one can perform on these correlators. Note first that these expressions are manifestly symmetric in $z \leftrightarrow \zb$ as well as $z,\zb \leftrightarrow \frac1z,\frac1\zb$. 
Next we can check some of the CFT data contained in this correlator, focusing on half-BPS operators that appear in the bulk and defect OPEs.
Since these are the leading operators in each $R$-symmetry channel, we can extract their contributions using non-supersymmetric conformal blocks \eqref{eq:defect-block}-\eqref{eq:R-sym-block}.
After extracting the relevant CFT data, we can then compare the results to supersymmetric localization. 
On the defect side, we predict the two following coefficients:
\begin{equation}
b_{21} b_{31} = \frac{3\sqrt{3}}{4} \frac{\sqrt{\lambda}}{N^2} + \ldots \,, \qquad 
b_{2(2)} b_{3(2)} = - \frac{\sqrt{3}}{2} \frac{\sqrt{\lambda}}{N^2}  + \ldots \,.
\end{equation}
The first one can be checked against equation (4.44) of \cite{Giombi:2018hsx}, and is found to match perfectly, while the second one is a new prediction.\footnote{In principle one could also compare this coefficient to localization using the methods of \cite{Giombi:2018qox,Giombi:2018hsx}. We reserve this analysis for future work.} On the bulk side, our result predicts the following coefficient
\begin{equation}
 a_{(2,3)} \lambda_{23(2,3)} 
 = \frac{\sqrt{3} }{16} \frac{\lambda}{N^2} 
 - \frac{15\sqrt{3}}{32} \frac{\sqrt{\lambda}}{N^2}
 + \ldots \, ,
\end{equation}
Using the matrix model methods of \cite{Billo:2018oog,Beccaria:2020ykg}, it is easy to determine the following formulae for arbitrary $J_1, J_2$ at large $N$:
\begin{equation}
a_{(J_1, J_2)} = \frac{\sqrt{J_1 J_2} \lambda}{2^{\frac{J_1 + J_2}{2}+2} N^2} \frac{I_{J_1 + J_2 - 1} (\sqrt{\lambda})}{I_1 (\sqrt{\lambda})}\,, \qquad \lambda_{J_1 J_2 (J_1, J_2)} = 1\,,
\label{eq:localization}
\end{equation}
and we observe perfect agreement with our result when expanding at strong coupling.

Let us stress that our correlator contains an infinite amount of CFT data, but here we only considered the values that can in principle be computed using localization.
A more detailed study of unprotected CFT data using the superconformal block expansion is relegated to future work.

\subsection{A closed form for the highest \texorpdfstring{$R$}{R}-symmetry channel}
\label{subsec:F0}

Thanks to the efficiency of the dispersion relation, we can obtain the correlation functions for high values of $J_1$ and $J_2$.
This allowed us to guess a closed formula for the channel $F_0$ for arbitrary external operators, which reads
\begin{align}
 F_0 (z,\zb) 
 & = \frac{\sqrt{J_1 J_2}}{2^{(|J_{12}|+2)/2} |J_{12}|} \frac{\sqrt{\lambda}}{N^2}
 \left( \frac{\sqrt{z\zb}}{(1-z)(1-\zb)} \right)^{\min(J_1,J_2)} 
 \left\lbrace 
   |J_{12}|^2 (1-\delta_{|J_{12}|,1})
  \phantom{\frac{1}{1}} \right. \notag \\*
& \left. \phantom{\frac{1}{1}} 
  - (|J_{12}|+1) \frac{(1-z)(1-\zb)}{(1-z\zb)^2} \bigg[ 
    (|J_{12}|+2)\, _2F_1 \Big( 1, |J_{12}|/2; |J_{12}|+2; 1-z\zb \Big) \right. \notag \\*
& \left.  \phantom{\frac{1}{1}} 
  + \frac12 |J_{12}| (z \zb - 3) - 2 \bigg] \right\rbrace\,.
  \label{eq:conjecture-F0}
\end{align}
There are several interesting features about this formula.
First, the limit of equal operators  $J_{12} \to 0$ is well defined, and if taken with care, it reproduces our previous conjecture \cite{Barrat:2021yvp}. 
Second, for any integer value of $J_{12}$, the hypergeometric function can be rewritten as a rational function of the cross-ratios and $\log z \zb$.
In particular, when $J_{12}$ is even the expression contains logarithms, while for $J_{12}$ odd the logarithms are absent.
We believe that is is also a feature of the full correlator $\Fm(z,\zb,\sigma)$.

Finally, notice that the case $J_{12} = \pm 1$ is somewhat special, because then the operator $\Om_1 \propto \text{tr}\, \phi$ can contribute to the discontinuity \eqref{eq:discblocks}. 
Here we have assumed gauge group $SU(N)$, in which case one should set $\Om_1 = 0$.
If we instead assume gauge group $U(N)$, then the operator $\Om_1$ contributes to the discontinuity, and the result is obtained from \eqref{eq:conjecture-F0} setting the Kronecker delta to zero $\delta_{|J_{12}|,1} \to 0$. 
If in the $U(N)$ case we restrict ourselves to $J_1 \ge J_2$ so that we can ignore absolute values, then the expression has the nice property of being \emph{analytic} in complex $J_1$ and $J_2$.

\section{Conclusions}
\label{sec:conclusions}

The main result of this work is the dispersion relation for defect CFT \eqref{eq:disp-rel}, that expresses two-point functions as an integral of a discontinuity times a kinematically fixed kernel. The formula is valid for general defects of codimension two or higher, and was checked against known results for monodromy defects and the Wilson line in $\Nm=4$ SYM. In the latter case, we also obtained new results for operators with unequal scaling dimension.
Thanks to the simplicity of the dispersion relation, these strong-coupling correlators can be computed very efficiently. 
This allowed us to find a closed form expression \eqref{eq:conjecture-F0} for the highest $R$-symmetry channel in the correlator of two arbitrary half-BPS operators.

As discussed in the introduction, it is possible to write a second dispersion relation in terms of $\dDisc F$. In principle one should be able to translate the four-point functions results from \cite{Carmi:2019cub} to the defect setup using the dictionary presented in \cite{Liendo:2019jpu}. This dictionary requires knowledge of four-point functions with unequal external operators, and although this case was not considered in \cite{Carmi:2019cub}, it appeared recently in \cite{Trinh:2021mll}.

The formulation of analytic bootstrap constraints in terms of dispersion relations is a promising approach, that elucidated the connection between several alternative techniques \cite{Caron-Huot:2020adz}, in particular Mellin space and the exact analytic functionals. It would be interesting to pursue an analog line of research in the defect setup. 
Regarding analytic functionals, it is in principle straightforward to obtain their kernels by expanding our dispersion relation in conformal blocks, see analog examples in \cite{Caron-Huot:2020adz,Giombi:2020xah}.
On the other hand, the Mellin space formalism for defect CFT was initiated in \cite{Rastelli:2017ecj,Goncalves:2018fwx}, but much remains to be done to show if it can be used in bootstrap calculations, or whether it is equivalent to our present methods. 

Mellin space played a crucial in the conjectured formula for half-BPS operators in $\Nm=4$ SYM at strong coupling \cite{Rastelli:2016nze}. Similar results for $\Nm=4$ SYM in the presence of a Wilson line were reported in \cite{Barrat:2021yvp} and further generalized in this work. A well defined question is how to translate these defect correlators to Mellin space, where we expect the structure of the functions to simplify.

\acknowledgments
We thank L. ~Bianchi, G.~Bliard, D.~Bonomi, S.~Dowker, V.~Forini, E.~Lauria, G.~Peveri, J.~Plefka, D.~Poland, P.~van Vliet for discussions and comments.
AG is particularly grateful to A.~Kaviraj for discussions on analytic functionals in BCFT.
JB’s research is funded by the Deutsche Forschungsgemeinschaft (DFG, German Research Foundation) - Projektnummer 417533893/ GRK2575 “Rethinking Quantum Field Theory”.
AG and PL acknowledge support from the DFG through the Emmy Noether research group ``The Conformal Bootstrap Program'' project number 400570283, and through the German-Israeli Project Cooperation (DIP) grant ”Holography and the Swampland”.

\appendix

\section{More correlators for the Wilson line defect in \texorpdfstring{$\Nm = 4$}{N=4} SYM}
\label{sec:appendixA}

In this appendix we provide more examples of correlators $\vvev{\Om_{J_1} \Om_{J_2}}$ for the two-point functions in $\Nm=4$ SYM with a Wilson line defect presented in section \ref{sec:newresults}.

\subsection{\texorpdfstring{$(J_1, J_2) = (2,4)$}{(J1, J2) = (2, 4)}}

We start by studying the case $(J_1, J_2) = (2,4)$. Following the same method as in section \ref{subsec:case23}, we find the following correlator:
\begin{align}
F_0 = & \frac{\sqrt{\lambda}}{ 2 \sqrt{2} N^2} \frac{z \zb}{(1-z)^2 (1-\zb)^2 (1-z\zb)^5} \left( 1+3(z+\zb) + z\zb(1-24(z+\zb)) + 64(z\zb)^2 \phantom{\frac{1}{1}} \right. \notag \\ 
& \left. -64(z \zb)^3 \left( 1- \frac{3}{8} (z+\zb) \right) - (z\zb)^4 (1+3(z+\zb)) - (z\zb)^5 \right. \notag \\
& \left.  \phantom{\frac{1}{1}} + 36 (z\zb)^2(1-z)(1-\zb) \log z\zb \right)\,, \\
F_1 = & \frac{\sqrt{\lambda}}{2\sqrt{2} N^2} \frac{z \zb}{(1-z)(1-\zb)(1-z\zb)^4} \left( 5 + 7(z+\zb) - 41 z\zb \left( 1 - \frac{22}{41} (z+\zb) \right) \right. \notag \\
& \left. - 41(z\zb)^2 \left(1 - \frac{7}{41} (z+\zb) \right) + 5(z\zb)^3 + 2 \frac{z\zb}{(1-z\zb)} \left( 12(z+\zb) - 38 z \zb \left( 1 - \frac{1}{19}(z+\zb) \right) \right. \right. \notag \\
& \left. \left. + 5(z\zb)^2 (1+z+\zb) - 4 (z\zb)^3 \left( 1 + \frac{1}{4}(z+\zb) \right) + (z\zb)^4 \right) \log z\zb \right) \notag \\
& +\frac{\sqrt{2} \sqrt{\lambda}}{N^2} \log (1+\sqrt{z\zb})\,, \\
F_2 = & - \frac{\sqrt{\lambda}}{8\sqrt{2} N^2} \frac{1}{\sqrt{z\zb} (1-z\zb)^4} \left( 4(z+\zb) + 9 \sqrt{z \zb} \left( 1 - \frac{2}{9} (z+\zb) \right) -16 z\zb (z+\zb) \right. \notag \\
& \left. - 58(z\zb)^{3/2} \left( 1 - \frac{10}{29}(z+\zb) \right) + 24(z\zb)^2(z+\zb) + 26 (z\zb)^{5/2} \left( 1 + \frac{10}{13} (z+\zb) \right) \right. \notag \\
& \left. - 58(z\zb)^{7/2} \left( 1 + \frac{1}{29} (z+\zb) \right) - 16(z\zb)^3 (z+\zb) + 9 (z\zb)^{9/2} + 4 (z\zb)^4 (z+\zb) \right. \notag \\
& \left. - \frac{4 (z\zb)^{5/2}}{1-z\zb} \left( 15 - 10(z+\zb) - 5 z\zb \left( 1 - \frac{1}{2}(z+\zb) \right) + 10(z\zb)^2 \left( 1 - \frac{1}{5} (z+\zb) \right) \right. \right. \notag \\
& \left. \left. - 2 (z\zb)^3 \left( 1 - \frac{1}{4}(z+\zb) \right) \right) \log z\zb \right) - \frac{\sqrt{\lambda}}{2\sqrt{2} N^2} \frac{4 z\zb-(1+z\zb)(z+\zb)}{z\zb} \log (1+\sqrt{z\zb})\,. &
\end{align}
To obtain the result, we used the following input on the bulk side
\begin{equation}
a_2 \lambda_{242} = \frac{\sqrt{2} \sqrt{\lambda}}{N^2} + \ldots \,, \qquad 
a_4 \lambda_{244} = \frac{\sqrt{2} \sqrt{\lambda}}{N^2} + \ldots \,,
\end{equation}
and the OPE coefficient for the defect identity, which cannot be probed by our bootstrap methods
\begin{equation}
a_2 a_4 
= \frac{1}{8 \sqrt{2}} \frac{\lambda }{N^2} 
- \frac{9}{8\sqrt{2}} \frac{\sqrt{\lambda}}{N^2} + \ldots \,.
\end{equation}

We can now extract the OPE coefficients of other half-BPS operators, and compare them to localization results. On the defect side, the output reads
\begin{equation}
b_{21} b_{41} = \frac{\sqrt{2} \sqrt{\lambda}}{N^2} + \ldots  \,, \qquad 
b_{2(2)} b_{4(2)} = \frac{1}{2\sqrt{2}} \frac{\sqrt{\lambda}}{N^2} + \ldots \,.
\end{equation}
The first coefficient is in perfect agreement with equation (2.31) in \cite{Barrat:2021yvp}, while the second has not been previously computed. On the bulk side we find that
\begin{equation}
a_{(2,4)} \lambda_{24(2,4)} 
= \frac{1}{8 \sqrt{2}} \frac{\lambda }{N^2} 
- \frac{3}{2\sqrt{2}} \frac{\sqrt{\lambda}}{N^2} + \ldots \, .
\end{equation}
This result is also in agreement with equation \eqref{eq:localization}.

\subsection{\texorpdfstring{$(J_1, J_2) = (3,4)$}{(J1, J2) = (3, 4)}}

We finally turn our attention to $(J_1, J_2) = (3,4)$, another correlator for which $J_2 - J_1 = 1$. 
As in the main text, we choose to work with $SU(N)$ gauge group, so we do not consider the contribution of $\Om_1$ to the discontinuity.
In this example there are four $R$-symmetry channels, and as for the other correlators, they can be fixed unambiguously using the method of section \ref{subsec:case23}. Here spin $s=0$ ambiguities are added the channel $F_2$, and spin $s=0,1$ ambiguities are added to $F_3$. We find the following result:
\begin{align}
F_0 = & - \frac{\sqrt{3} \sqrt{\lambda}}{\sqrt{2} N^2} \frac{(z \zb)^{3/2} (1+6\sqrt{z \zb}+10 z \zb + 6 (z\zb)^{3/2} + (z\zb)^2)}{(1-z)^2(1-\zb)^2(1+\sqrt{z\zb})^6}\,,  \\
F_1 = & \frac{\sqrt{3} \sqrt{\lambda}}{\sqrt{2} N^2} \frac{(z\zb)^{3/2}}{(1-z)^2(1-\zb)^2(1+\sqrt{z\zb})^6} \left( 1 + 5(z+\zb) +6 \sqrt{z\zb} (1+3(z+\zb)) \phantom{\frac{1}{1}} \right. \notag \\
& \left. + 26 z\zb \left( 1 + \frac{5}{26} (z+\zb) \right) +6(z\zb)^{3/2} + (z\zb)^2 \right)\,, \\
F_2 = & \frac{\sqrt{3} \sqrt{\lambda}}{2\sqrt{2} N^2} \frac{\sqrt{z\zb}}{(1-z)(1-\zb)(1-z\zb)^6 (1+\sqrt{z\zb})^4} \left( 40 z\zb (1-z\zb)^6 \phantom{\sqrt{z\zb}} \right. \notag \\
& \left. + (1-z)(1-\zb)(1+\sqrt{z\zb})^4 \left( 3 - 49 z \zb + 144(z\zb)^{3/2} - 210 (z\zb)^2 + 224 (z\zb)^{5/2} \right. \right. \notag \\
& \left. \left. - 210 (z\zb)^3 + 144(z\zb)^{7/2} - 49(z\zb)^4 + 3 (z\zb)^5 \right) \right)\,,  \\
F_3 = & \frac{23 \sqrt{3} \sqrt{\lambda}}{32 \sqrt{2} N^2} \left( -1 + \frac{16}{23 (1-z\zb)^6} \left( - \sqrt{z\zb} (3-(z+\zb)) + 41 (z\zb)^{3/2} \left( 1 - \frac{20}{41} (z+\zb) \right) \right. \right. \notag \\
& \left. \left. - 96 (z\zb)^2 \left( 1 - \frac{2}{3} (z+\zb) \right) + 90 (z\zb)^{5/2} \left( 1 - \frac{32}{45} (z\zb)^{1/2} \right) (1-(z+\zb)) \right. \right. \notag \\
& \left. \left. + 90 (z\zb)^{7/2} \left( 1 - \frac{2}{9} (z+\zb) \right) - 96 (z\zb)^4 + (z\zb)^{9/2} (41 + (z+\zb)) - 3 (z\zb)^{11/2} \right) \right)\,.
\end{align}

To obtain this result, we used as an input the following OPE coefficients on the bulk side
\begin{equation}
\lambda_{343} a_3 = \frac{3 \sqrt{3} \sqrt{\lambda}}{2 \sqrt{2} N^2} + \ldots \,, \qquad 
\lambda_{345} a_5 = \frac{5 \sqrt{3} \sqrt{\lambda}}{4 \sqrt{2} N^2} + \ldots \,,
\end{equation}
while for the defect identity we used
\begin{equation}
a_3 a_4  
= \frac{1}{16} \sqrt{\frac{3}{2}} \frac{\lambda}{N^2}
- \frac{23 \sqrt{3}}{32 \sqrt{2}} \frac{\sqrt{\lambda}}{N^2} + \ldots \,.
\end{equation}

As for the other examples, we can check that these correlators lead to the correct CFT data for half-BPS operators that appear in the OPEs. On the defect side, leaving the input apart, we have:
\begin{equation}
b_{31} b_{41} = \frac{3}{2} \sqrt{\frac{3}{2}} \frac{\sqrt{\lambda}}{N^2} + \ldots \,, \quad 
b_{32} b_{42} = 0 + \ldots \,, \quad 
b_{3(3)} b_{4(3)} = - \sqrt{\frac{3}{2}} \frac{\sqrt{\lambda}}{N^2} + \ldots \,.
\end{equation}
The two first coefficients match equation (2.31) in \cite{Barrat:2021yvp}, while the third one is a new prediction.
On the bulk side the new OPE coefficient is
\begin{equation}
a_{(3,4)} \lambda_{34(3,4)} 
= \frac{1}{16} \sqrt{\frac{3}{2}} \frac{\lambda}{N^2}
-\frac{35}{32} \sqrt{\frac{3}{2}} \frac{\sqrt{\lambda}}{N^2} + \ldots \,,
\end{equation}
which matches the localization prediction given in \eqref{eq:localization}.

\bibliography{./auxi/biblio}
\bibliographystyle{./auxi/JHEP}

\end{document}